
\magnification=\magstep1          \def\sp{\,\,\,} \overfullrule=0pt
  \def\c{\chi}
 \def\p{{\cal P}} \def\eps{\epsilon}
\def\la{\lambda}   \def\R{{\cal K}} \def\Z{{\bf Z}} \def\J{{\cal J}}
\def\equi{\,{\buildrel \rm def \over =}\,} \def\C{{\cal C}}
\def\eg{{\it e.g.}$\sp$} \def\ie{{\it i.e.}$\sp$}  \def\O{{\cal O}}
 \font\huge=cmr10 scaled \magstep2
\font\small=cmr7

 \def\COMM{[8]} \def\CG{[7]} \def\ALZ{[1]} \def\BER{[4]}
\def\BMW{[3]} \def\KAC{[15]} \def\MS{[18]} \def\CR{[6]} \def\BI{[2]}
\def\GH{[11]} \def\SU{[9]}  \def\AA{[10]} \def\RTW{[21]}
\def\SCH{[17]} \def\ARSY{[22]} \def\KR{[16]} \def\RUE{[19]}
\def\STA{[24]} \def\PF{[14]} \def\GW{[13]} \def\RU{[20]}  \def\ST{[25]}
\def\HET{[12]} \def\CIZ{[5]} \def\MER{[23]}

{ \nopagenumbers
\rightline{April, 1994}
\bigskip \bigskip
\centerline{{\bf \huge The Classification of SU(3)}}
\bigskip
\centerline{{\bf \huge Modular Invariants Revisited}}
\bigskip \bigskip\bigskip
\centerline{{Terry Gannon\footnote{*}{\small email: gannon@ihes.fr}}}
\centerline{{\it Institut des Hautes Etudes Scientifiques}}
\centerline{{\it 91440 Bures-sur-Yvette, France}}
\bigskip \bigskip \centerline{\bf Abstract}\bigskip
The SU(3) modular invariant partition functions were first completely
classified in Ref.\ \SU. The purpose of these notes is four-fold:
\item{(i)} Here we accomplish the SU(3) classification using only
the most basic facts: modular invariance; $M_{\la\mu}\in{\bf Z}_{\ge}$; and
$M_{00}=1$. In \SU{} we made use of less elementary results from
Moore-Seiberg, in addition to these 3 basic facts.
\item{(ii)} Ref.\ \SU{}  was completed well over a year ago. Since then
I have found a number of significant simplifications to the general argument.
They are all included here.
\item{(iii)} A number
of people have complained that some of the arguments in \SU{} were hard to
follow. I have tried here to be as explicit and as clear as possible.
\item{(iv)} Hidden in \SU{} were a number of smaller results which should
be of independent value. These are explicitly mentioned here.
\vfill \eject} \pageno=1

\noindent{{\bf 1. Introduction}} \bigskip

This paper focuses exclusively on the classification of SU(3) WZNW partition
functions, though many of
the results and techniques work in much greater generality. See \SU{}
for the motivation for the problem.

A level $k$ {\it partition function} looks like
$$Z(\tau,z_L,z_R)=\sum_{\la,\mu\in\p^k} M_{\la\mu} \,\c_{\la}^{k}(\tau,z_L,0)
\,\c_{\mu}^{k}(\tau,z_R,0)^*. \eqno(1.1)$$
$\c_{\lambda}^{k}$
 is the {\it normalized character} \KAC{} of the representation of the
affine Kac-Moody algebra $A_2^{(1)}$ with (horizontal) highest weight $\la$
and level
$k$; it can be thought of as a complex-valued function of the Cartan
subalgebra of $A_2^{(1)}$,
\ie a function of a complex 2-vector $z$ and complex numbers $\tau,u$ (we
will always take $u=0$).
The (finite) sum in eq.(1.1) is over  the horizontal highest weights
$\la,\mu\in\p^k$ of level $k$:
$$\p^k=P^{k+3}_{++}\equi  \{(\la_1,\la_2)\,|\,\la_1,\la_2\in\Z,\sp
0< \la_1,\la_2,\sp \la_1+\la_2< k+3\}.\eqno(1.2)$$
We will always identify a weight $\la$ with its Dynkin labels $\la_1,\la_2$.
The quantity $k+3$ will appear throughout. It is called the {\it height}
and will be denoted $n$.

\medskip\noindent{{\bf Warning 1}}\quad These weights $\la$ are only {\it
pairs} $(\la_1,\la_2)$, not {\it triples} $(\la_0,\la_1,\la_2)$. That is,
they are the horizontal projections of the affine weights: $\la_0$ is
dropped, since it is redundant once the level is known. Also, these weights are
shifted by $\rho=(1,1)$, so the `identity', or `vacuum', corresponds
to $\la=\rho$, not $\la=(0,0)$. For some purposes it would have been
more convenient in this paper to not have shifted by $\rho$, but for
most purposes this convention is better.

\medskip\noindent{{\bf Warning 2}}\quad The characters $\chi$ in this
paper will depend on a complex 2-vector $z$. Many people working on these
problems use `restricted characters', \ie put $z=0$. I believe
this is a mistake, that the math really demands that $z$ be included.

\medskip\noindent{{\bf Warning 3}}\quad Reluctantly, I have decided to
make some notation changes from the \SU{} paper. For example, to avoid
confusion with the fusion coefficients I'll let $M_{\la\mu}$ and not
$N_{\la\mu}$ denote the matrix in (1.1). Also, I'll succumb to standard
convention and let $n$, not $k'$, denote the height $k+3$. \medskip

Because (and only because) we have $z_L,z_R\ne 0$, there is a one-to-one
correspondence between the partition function $Z$ and its coefficient
matrix $M$. We will freely identify them.

The characters $\c^k_\la$ for fixed $k$ define a unitary representation
of the modular group $SL(2,\Z)$. In particular:
$$\eqalignno{\c_\la^k(\tau+1,z)&=\sum_{\mu\in \p^k}\big(T^{(n)}
\big)_{\la\mu}\,\c_{\mu}^{k}(\tau,z),\sp \sp{\rm where}&(1.3a)\cr
\big(T^{(n)}\big)_{\la\mu}&=
e_n(-\la_1^2-\la_1\la_2-\la_2^2+n)\,\delta_{\la\mu};&(1.3b)\cr
\c_\la^{k}(-1/\tau,z/\tau)&=\exp[k\pi iz^2/\tau]\,\sum_{\mu\in \p^k}
\big(S^{(n)}\big)_{\la\mu}\c_{\mu}^{k}(\tau,z),\sp \sp{\rm where}&(1.3c)\cr
\big(S^{(n)}\big)_{\la\mu}=
&{-i\over \sqrt{3}n}\bigr\{ e_n(2\la_1\mu_1+\la_1\mu_2+\la_2\mu_1+
2\la_2\mu_2)+e_n(\la_2\mu_1-\la_1\mu_1-2\la_1\mu_2-\la_2\mu_2)
&\cr & +e_n(\la_1\mu_2-\la_1\mu_1-2\la_2\mu_1-\la_2\mu_2)-e_n(-2\la_1\mu_2
-\la_1\mu_1-\la_2\mu_2-2\la_2\mu_1)&\cr &
-e_n(2\la_1\mu_1+\la_1\mu_2+\la_2\mu_1-\la_2\mu_2)-e_n(\la_1\mu_2-\la_1\mu_1+
\la_2\mu_1+2\la_2\mu_2)\}&(1.3d)\cr}$$
where $e_n(x)\equi \exp[{-2\pi i x\over 3n}]$.
The matrices $T^{(n)}$ and $S^{(n)}$ are unitary and symmetric.

Our task will be to find all $Z$ in (1.1) satisfying the following 3
properties:\medskip
\item{(P1)} {\it modular invariance}. This is equivalent to
the matrix equations:
$$\eqalignno{T^{(n)\dag}\,M\,T^{(n)}=M,&\quad i.e.\quad M\,T^{(n)}=T^{(n)}\,M,
 &(1.4a)\cr
S^{(n)\dag}\,M\,S^{(n)}=M,&\quad i.e.\quad M\,S^{(n)}=S^{(n)}\,M; &(1.4b)\cr}$$

\item{(P2)} the coefficients
$M_{\la \mu}$ in eq.(1.1) must be non-negative integers; and

\item{(P3)} we must have $M_{\rho\rho}=1$, where $\rho=(1,1)$.\medskip

We will call any modular invariant function $Z$ of the form (1.1),
an {\it invariant}. $Z$ will be called {\it positive} if
in addition each $M_{\la\mu}\ge 0$, and {\it physical}
 if it satisfies (P1), (P2), and (P3).
Our task is to find all physical invariants for each level $k$.
There are other properties a physically reasonable
partition function should satisfy, but for a number of reasons it is
preferable to limit attention to as small a number of properties as possible.
{\it In this paper, only (P1)-(P3) will be used}.

The $SU(3)$ classification has had a fairly long history. Ref.\ \BI{} tried
to understand the space of all invariants, for
any $SU(N)_k$; although this approach works for $SU(2)_k$, it was too messy
even for $SU(3)_k$. But this work was used by \RTW{} to prove the $SU(3)_k$
classification for $k+3$ prime. It also led to the {\it parity rule},
which turned out to be so important in the $SU(3)_k$ classification -- this
was independently discovered in \COMM{} and \ARSY. In work done simultaneously
but independently of \SU, Ref.\ \RUE{} classified the
{\it permutation invariants} of $SU(3)_k$ (see eq.$(2.2a)$ below);
his argument is much longer than
the one given in \SU{} but has the advantage of not requiring fusion rules,
so it should generalize more easily to other $SU(N)_k$. \ARSY{} used it, and an
amazing coincidence with the Fermat curves \KR, to prove the $SU(3)_k$
classification for $k+3$ coprime to 6, and for $k+3=2^i$ and $k+3=3^i$.
\STA{} used the Knizhnik-Zamolodchikov equations to find all local
extensions of the $SU(3)$ chiral algebra; in our language this gives
the possible $\rho$-couplings (see Sect.4 below).
But the first, and to my knowledge the
only, classification of $SU(3)_k$ physical invariants for all $k$ was given
in \SU. It was done independent of -- in fact oblivious to -- all the above
work, apart from \BI{} (and the $SU(3)_k$ fusion rules, calculated in \BMW).

Within this context, the only thing this present paper really adds is that
it accomplishes the classification using only the properties (P1)-(P3). But
it also simplifies and expands out more explicitly
the arguments in \SU, rewriting some completely. This should make the
whole argument much more accessible.
It also makes explicit some  results hidden inside \SU; these should be
useful in other classifications -- \eg Claims 1 and 2 and Prop.1 are used
in \GW.

The 6 outer automorphisms of $A_2^{(1)}$ are generated by
$C$ (order 2) and $A$ (order 3). These act on
the horizontal weights $(\la_1,\la_2)\in \p^k$ in this way:
$$\eqalignno{C(\la_1,\la_2)&=(\la_2,\la_1),&(1.5a)\cr
A(\la_1,\la_2)&=(n-\la_1-\la_2,\la_1).&(1.5b)\cr}$$
Note that $A^2(\la_1,\la_2)=(\la_2,n-\la_1-\la_2)$.
The $A^a$ are called {\it simple currents}, and $C$
is called the {\it (charge) conjugation}. They obey the relations
$$\eqalignno{T_{C\la,C\mu}^{(n)}=&\,T_{\la\mu}^{(n)},&(1.6a)\cr
S_{C\la,\mu}^{(n)}=&\,S_{\la\mu}^{(n)*},&(1.6b)\cr
T_{A^a\la,A^a\mu}^{(n)}=&\,\exp[2\pi i\{a^2n-a\,t(\la)\}/3]\,
T_{\la\mu}^{(n)},&(1.6c)\cr
S_{A^a\la,A^b\mu}^{(n)}=&\,\exp[2\pi i\{b\,t(\la)+a\,t(\mu)+
nab\}/3]\,S_{\la\mu}^{(n)},&(1.6d)\cr}$$
where $t(\la)=\la_1-\la_2$ is called the {\it triality} of $\la$.
Write $\O$ for this 6 element group, and $\O\la$ for the orbit of $\la$ under
$\O$. Write $\O_0$ for the 3 element subgroup generated by $A$.

Our goal is to prove that the only level $k$ SU(3) physical invariants are:
$$\eqalignno{{\cal A}_k\equi & \sum_{\la\in \p^k} |\c_\la^k|^2,\qquad
\forall k\ge 1;&(1.7a)\cr
{\cal D}_k\equi & \sum_{\la\in \p^k}\chi^k_{\la}\,
\chi^{k*}_{A^{k\,t(\la)}\la},
\quad {\rm \sp for \sp}k\not\equiv 0 \sp{\rm (mod\sp 3) \sp and
\sp}k\ge 4;&(1.7b)\cr
{\cal D}_k\equi & \,{1\over 3}
\sum_{{\la\in \p^k \atop t(\la)\equiv 0 \sp({\rm mod}\sp 3)}}
\,|\chi^k_{\la}+\chi_{A\la}^k+\chi^k_{A^2\la}|^2,
{\rm \sp for \sp}k\equiv 0 \sp{\rm (mod\sp 3)};&(1.7c)\cr
{\cal E}_5\equi & |\c^5_{1,1}+\c^5_{3,3}|^2+|\c^5_{1,3}+\c^5_{4,3}|^2+
|\c^5_{3,1}+\c^5_{3,4}|^2&\cr & +|\c^5_{3,2}+\c^5_{1,6}|^2+|\c^5_{4,1}+
\c^5_{1,4}|^2+|\c^5_{2,3}+\c^5_{6,1}|^2;&(1.7d)\cr
{\cal E}_9^{(1)}\equi &|\c^9_{1,1}+\c^9_{1,10}+\c^9_{10,1}+\c^9_{5,5}+
\c^9_{5,2}+\c^9_{2,5}|^2+2|\c^9_{3,3}+\c^9_{3,6}+\c^9_{6,3}|^2;&(1.7e)
\cr {\cal E}_9^{(2)}\equi &|\c^9_{1,1}+\c^9_{10,1}+\c^9_{1,10}|^2+|\c^9_{3,3}+
\c^9_{3,6}+\c^9_{6,3}|^2+2|\c^9_{4,4}|^2& \cr &+|\c^9_{1,4}+\c^9_{7,1}+
\c^9_{4,7}|^2+|\c^9_{4,1}+\c^9_{1,7}+\c^9_{7,4}|^2+|\c^9_{5,5}+\c^9_{5,2}+
\c^9_{2,5}|^2&\cr &+(\c^9_{2,2}+\c^9_{2,8}+\c^9_{8,2})\c^{9*}_{4,4}+
\c^9_{4,4}(\c^{9*}_{2,2}+\c^{9*}_{2,8}+\c^{9*}_{8,2});&(1.7f)\cr
{\cal E}_{21}\equi & |\c^{21}_{1,1}+\c^{21}_{5,5}+\c^{21}_{7,7}+
\c^{21}_{11,11}+\c^{21}_{22,1}+\c^{21}_{1,22}&\cr &+\c^{21}_{14,5}+
\c^{21}_{5,14}+\c^{21}_{11,2}+\c^{21}_{2,11}+\c^{21}_{10,7}+\c^{21}_{7,10}
|^2&\cr &+|\c^{21}_{16,7}+\c^{21}_{7,16}+\c^{21}_{16,1}+\c^{21}_{1,16}+
\c^{21}_{11,8}+\c^{21}_{8,11} &\cr &+\c^{21}_{11,5}+\c^{21}_{5,11}+
\c^{21}_{8,5}+\c^{21}_{5,8}+\c^{21}_{7,1}+\c^{21}_{1,7}|^2;&(1.7g)\cr}$$
together with their {\it conjugations} $Z^c$ under $C$, defined by:
$$\bigl( M^c\bigr)_{\la\mu}=M_{C\la,\mu}.\eqno(1.7h)$$
Note that ${\cal D}_3={\cal D}_3^c$, ${\cal D}_6={\cal D}_6^c$,
${\cal E}_9^{(1)}={\cal E}_9^{(1)c}$, and ${\cal E}_{21}={\cal E}_{21}^c$.

The invariants $(1.7b)$ were first found in \ALZ, while $(1.7c)$ was found
in \BER. The exceptionals $(1.7d,e,g)$ were found in \CR, while $(1.7f$)
was found in \MS.

The remainder of this paper is devoted toward proving that eqs.(1.7) exhaust
all SU(3) physical invariants.\smallskip

\item{{\it Sec.{} 2}} We state here the tools which will be used
to accomplish this, some of which are new to \SU.

\item{{\it Sec.{} 3}} We find all permutation invariants for each level $k$.
This argument is based on Sect.3 of \SU, but has been improved in a
number of places.\smallskip

\item{{\it Sec.{} 4}} For each $k$, we use the `parity rule' to find all
weights $\la\in \p^k$ which can `couple to $\rho$'. This is the most difficult
part of the paper; it is based on Sect. 4 of \SU, but the arguments are given
in more detail here, and the most complicated case in \SU{} (namely
$n\equiv 2$ (mod 4)) has been completely rewritten.
The arguments in this section are elementary but tedious and involve
investigating several cases.\smallskip

\item{{\it Sec.{} 5}} Everything is put together here.
This section is completely rewritten and considerably simplified from
Sect. 5 of \SU. One of our tasks here is to find all automorphisms of the
`simple current extension' when 3 divides $k$, but this argument
essentially reduces to the one in Sect. 3.

\item{{\it Sec.{} 6}} This new section explicitly
handles the four anomolous levels $k=5,9,21,57$.

\bigskip \bigskip \noindent{{\bf 2. The parity rule and other tools}}
\bigskip

In this section we collect together the various tools(=lemmas) we will
be using. All these apply in a far greater context than merely SU(3),
but we will state and prove only what we need.

The weight $\rho$ is very special. For one thing, there is the important
property that
$$\eqalignno{S^{(n)}_{(a,a),\la}=&{8\over \sqrt{3}n}\sin({\pi a\la_1\over n})\,
\sin({\pi a\la_2\over n}) \,\sin({\pi a\,(\la_1+\la_2)\over n}),\quad {\rm
for\ all}\ 1\le a\le {n-1\over 2}, &(2.1a)\cr
{\rm hence}&\qquad S^{(n)}_{\rho\la}\ge S^{(n)}_{\rho\rho}>0.&(2.1b)\cr}$$
When $a=1$, equality holds in $(2.1b)$ iff $\la\in\O_0\rho$. $(2.1b$) and (P3)
together suggest the possibility that the values
$M_{\rho\mu}$, $M_{\la\rho}$ may be important. Indeed this is the case:
our first three lemmas given below all tell us {\it global} information
about $M$, given the {\it local} knowledge $M_{\rho\mu},M_{\la\rho}$.

Because $T^{(n)}$ is diagonal, $(1.4a$) is easy to solve: $M$ commutes with
$T^{(n)}$ iff
$$\la_1^2+\la_1\la_2+\la_2^2\equiv \mu_1^2+\mu_1\mu_2+\mu_2^2\qquad {\rm
whenever\ }M_{\la\mu}\ne 0.\eqno(2.1c)$$
A much harder task is to obtain useful information from $(1.4b$). This is
the purpose of this section.

First some definitions. Call a physical invariant $M$ a {\it permutation
invariant} if there exists a permutation $\sigma$ of $\p^k$ such that
$$M_{\la\mu}=\delta_{\mu,\sigma\la}.\eqno(2.2a)$$
For a given invariant $M$, let
$$\eqalignno{\R^\mu_L(M)=&\{\la\in\p^k\,|\,M_{\la\mu}\ne 0\},&(2.2b)\cr
\p_L(M)=&\{\la\in\p^k\,|\,\exists \mu\sp {\rm such \sp that}\sp M_{\la\mu}\ne
0\}=\bigcup_{\mu\in\p^k} \R^\mu_L(M),&(2.2c)\cr
\J_L(M)=&\{A^a\in \O_0\,|\, A^a\rho\in \R^\rho_L(M)\},&(2.2d)\cr
s^\la_L(M)=&\sum_{\mu\in\p^k}S^{(n)}_{\la\mu}\,M_{\mu\rho},&(2.2e)\cr}$$
and define $\R^\la_R(M)$, $\p_R(M)$, $\J_R(M)$ and $s^\mu_R(M)$ similarly
-- \eg $\R^\la_R(M)=\{\mu\,|\,M_{\la\mu}\ne 0\}$. Let $[\la]$ denote the
orbit $\J_L(M)\,\la$, and $[\mu]'$ denote the orbit $\J_R(M)\,\mu$.

\medskip\noindent{\bf Lemma 1.}\quad{\bf (a)} Let $M$ be any positive
invariant. For each $\la,\mu\in\p^k$, both $s^\la_L(M),s^\mu_R(M)\ge 0$.
Also, $s^\la_L(M)>0$ iff $\la\in\p_L(M)$; $s^\mu_R(M)>0$ iff $\mu\in\p_R(M)$.

\item{\bf (b)} Let $M$ be any physical invariant. Then $\J_L(M)$
and $\J_R(M)$ are groups, \ie equal either $\{A^0\}$ or $\O_0=\{A^0,A^1,A^2\}$.

\item{\bf (c)} Let $M$ be any physical invariant. Then $M_{\la\mu}
=M_{A^a\la,A^b\mu}$ for any $A^a\in\J_L(M)$, $A^b\in\J_R(M)$.
In other words, there exists a matrix $M^e$ whose entries are non-negative
integers,  such that $M^e_{[\rho],[\rho]'}=1$ and
$$\eqalign{Z=&\sum_{\la,\mu\in\p^k} M_{\la\mu}\,\chi^k_\la\,\chi_\mu^{k*}=
\sum_{[\la],[\mu]'
\subset\p^k} M^e_{[\la],[\mu]'}\,ch_{[\la]}\,ch'_{[\mu]'}{}^*,\cr {\rm where}&
\quad ch_{[\la]}=\sum_{\nu\in[\la]}\chi_\nu^k,\qquad ch'_{[\mu]'}=\sum_{\nu\in
[\mu]'}\chi_\nu^k.\cr}$$

\item{\bf (d)} Let $M$ be a physical invariant. If say $M_{\la\rho}\le
M_{\rho\mu}$ $\forall \la\in\p^k$, then $M_{\rho\la}=M_{\la\rho}$
$\forall \la\in\p^k$ and $\R^\rho_L(M)=\R^\rho_R(M)$.

\item{\bf (e)} Let $M$ be a physical invariant, and suppose $\R^\rho_L
(M)=[\rho]$, $\R^\rho_R(M)=[\rho]'$. Then the cardinalities $\|\J_L(M)\|=
\|\J_R(M)\|$ are equal, so $\J_L(M)=\J_R(M)$, and $\p_L(M)$ equals the set of
all weights with zero charge with respect to $\J_L(M)$, \ie
$$\p_L(M)=\{\la\in\p^k\,|\,a\,t(\la)\equiv 0\sp({\rm mod}\sp 3)\sp
{\rm whenever\sp}A^a\in\J_L(M)\}.$$

\noindent{\it Proof}\quad {\bf 1(a)} Evaluating $\bigl(S^{(n)}M\bigr)_{\la\rho}
=\bigl(MS^{(n)}\bigr)_{\la\rho}$ (see $(1.4b)$) gives us
$$s^\la_L(M)\equi\sum_{\mu\in\p^k}S^{(n)}_{\la\mu}\,M_{\mu\rho}=
\sum_{\mu\in\p^k}M_{\la\mu}\,S^{(n)}_{\mu\rho}.\eqno(2.3a)$$
The RHS of $(2.3a)$ is $\ge 0$, since each $S_{\rho\mu}^{(n)}>0$, by
$(2.1b)$, and each $M_{\mu\la}\ge 0$, by (P2). This gives us the first
part of 1(a). In fact the RHS of $(2.3a)$
will be $>0$ iff some $M_{\la\mu}>0$, \ie iff $\la\in\p_L(M)$. This gives us
the second.

{\bf 1(b)} From $M=S^{(n)\dag}MS^{(n)}$ and $(1.6b$) we get
$$\eqalignno{M_{A^a\rho,\rho}=\bigl|M_{A^a\rho,\rho}\bigr|&=\bigl|
\sum_{\la,\mu}S^{(n)*}_{A^a\rho,\la}\,M_{\la\mu}
\,S^{(n)}_{\mu\rho}\bigr|=\bigl|\sum_{\la,\mu} S^{(n)*}_{\rho\la}\,M_{\la\mu}\,
S^{(n)}_{\mu\rho}\,\exp[-2\pi i a\,t(\la)/3]\bigr|&\cr &\le \sum_{\la,\mu}
S^{(n)*}_{\rho\la}\,M_{\la\mu}\,S^{(n)}_{\mu\rho}=M_{\rho\rho}=1.&(2.3b)\cr}$$
In deriving this we also used (P2) and and eq.(2.1$b)$. Equality
will happen in $(2.3b)$ iff
$a\,t(\la)\equiv 0$ (mod 3) for all $\la\in \p_L(M)$, so 1(b) follows.

{\bf 1(c)} As in $1(b)$, we get for any $A^a\in\J_L(M)$, $A^b\in\J_R(M)$,
$$\eqalignno{M_{A^a\la,A^b\mu}&=\sum_{\la',\mu'}S^{(n)*}_{A^a\la,\la'}\,
M_{\la'\mu'}\,S^{(n)}_{\mu',A^b\mu}=\sum_{\la',\mu'} S^{(n)*}_{\la\la'}\,
M_{\la'\mu'}\,S^{(n)}_{\mu'\mu}\,\exp[2\pi i \{b\,t(\mu')-a\,t(\la')\}/3]&\cr
&= \sum_{\la',\mu'}S^{(n)*}_{\la\la'}\,M_{\la'\mu'}\,S^{(n)}_{\mu'\mu}
=M_{\la\mu}.&(2.3c)\cr}$$
The third equal sign appears in $(2.3c$) because we learned in the proof
of 1(b) that $A^a\in\J_L(M)$ iff, for all $\la'\in\p_L(M)$, $a\,t(\la')\equiv
0$ (mod 3).

{\bf 1(d)} By $(1.4b$) we get
$$\sum_{\la\in\p^k}
S^{(n)}_{\rho\la}\,M_{\la\rho}=\bigl(S^{(n)}M\bigr)_{\rho\rho}=\bigl(M\,S^{(n)}
\bigr)_{\rho\rho}=\sum_{\la\in\p^k}M_{\rho\la}\,S^{(n)}_{\la\rho}=
\sum_{\la\in\p^k}M_{\la\rho}\,S^{(n)}_{\la\rho}.\eqno(2.3d)$$
But by $(2.1b$), each $S^{(n)}_{\rho\la}=S^{(n)}_{\la\rho}>0$. Because equality
must hold in $(2.3d$), the desired conclusion holds.

{\bf 1(e)} Each $M_{\rho\mu},M_{\la\rho}\in\{0,1\}$, by 1(c). Thus
$\|\J_L(M)\|=\|\J_R(M)\|$ follows from 1(b),1(d). In the proof of
1(b) we found that $\la\in\p_L(M)$ iff $\la$ has
zero charge w.r.t. all $\J_L(M)$.\qquad QED\medskip

Lemma 1 was first proved in \AA. It will play an important role in Sect.\ 5.
Of these, 1(c) is the most important. The hypothesis in 1(d) is often
satisfied, because most $M$ have $M_{\rho\mu},M_{\la\rho}\in\{0,1\}$.
In the general case (\ie not SU(3)), we
still have $\|\J_L(M)\|=\|\J_R(M)\|$ in 1(e), but this no longer will
necessarily mean $\J_L(M)=\J_R(M)$.

The next lemma is Theorem 3 in \COMM. The proof was given there, and should
be simple and explicit enough that there is no need to repeat it here. The
only additional result here is that $\R^\rho_L(M)=\{\rho\}$ iff $\R^\rho_R(M)
=\{\rho\}$; this follows immediately from 1(d).

\medskip \noindent{\bf Lemma 2.}\quad Let $M$ be a physical invariant. Then
$\R^\rho_L(M)=\{\rho\}$ iff $\R^{\rho}_R(M)=\{\rho\}$ iff $M$ is a permutation
invariant (see (2.2$a$)).\medskip

Note from $(1.4)$ that the matrix product $MM'$ of two invariants
$M$ and $M'$ is again an invariant (at the same level). This should be an
important fact, and it is quite possible it has not been exploited enough.
Anyways, the next lemma is the main place this property is used.
It is proved using the Perron-Frobenius theory of
non-negative matrices \PF, and can be thought of loosely as a generalization
of Lemma 2. It will be used in Sect.\ 5 to significantly restrict the
possibilities for the sets $\R^\la_{L,R}(M)$ of a given physical invariant
$M$, given $\R^\rho_{L,R}(M)$, and also to bound the values of $M_{\la\mu}$.

Any matrix $M$ can be written as a direct sum
$$M=\bigoplus_{\ell=1}^\alpha B_\ell=\left(
\matrix{B_1&0&\cdots&0\cr 0&B_2&&0\cr
\vdots&&\ddots&\vdots\cr 0&0&\cdots&B_\alpha\cr}\right),\eqno(2.4a)$$
of {\it indecomposable} blocks $B_\ell$. In $(2.4a$) every index $i$ of $M$ is
`contained' in one and only one $B_\ell$.
By a {\it non-negative matrix} we mean a square matrix $M$ with non-negative
real entries. Any such matrix has a non-negative real
eigenvalue $r=r(M)$ (called the {\it Perron-Frobenius eigenvalue} \PF) with the
property that $r\ge |s|$ for all other (possibly complex) eigenvalues $s$
of $M$. The number $r(M)$ has many nice properties, for example:
$$\eqalignno{{\rm min}_i \sum_j M_{ij}\le &r(M)\le {\rm max}_i \sum_j M_{ij},
&(2.4b)\cr
{\rm max}_i M_{ii}\le &r(M);&(2.4c)\cr}$$
provided $M$ is indecomposable and symmetric, either equality holds in $(2.4b)$
iff each row sum $\sum_j M_{ij}$ is equal, and equality holds in (2.4$c$)
iff $M$ is the $1\times 1$ matrix $M=(r)$. Also, there is an eigenvector $v$
with eigenvalue $r$ with components $v_i\ge 0$.

For example, consider the $m\times m$ matrix
$$B_{(\ell,m)}=\left(\matrix{\ell&\cdots&\ell\cr \vdots&&\vdots\cr \ell&
\cdots&\ell\cr}\right).\eqno(2.4d)$$
Its eigenvalues are 0 (multiplicity $m-1$) and $m\ell$ (multiplicity 1).
Therefore $r(B_{(\ell,m)})=m\ell$. The eigenvector $v$ here is $v=(1,\ldots,
1)$. These matrices have the important property that they are proportional
to their square. They occur frequently in modular invariants.

\medskip\noindent{\bf Lemma 3.}\quad{\bf (a)} Let $M$ be a positive invariant
with non-zero indecomposable blocks $B_\ell$.
Call $B_1$ the block `containing' $\rho$. Then $r(B_\ell)\le r(B_1)$ for all
$\ell$.

\item{\bf (b)} Suppose now that $B_1^2=rB_1$, for some scalar
$r$. Then $r(B_\ell)=r$ for all $\ell$. If in addition $B^T_1=B_1$, then
each $M_{\la\mu}\le r$.

\item{\bf (c)} Now let $M$ be a physical invariant, and
 $\R^\rho_L(M)=[\rho]$, $\R^\rho_R(M)=[\rho]'$. Suppose $M_{\la\mu}\ne 0$,
where $\la$ is not a fixed point of $\J_L(M)$ (\ie $J\in\J_L(M)$ and $J\la
=\la$ implies
$J=A^0$) and $\mu$ is not a fixed point of $\J_R(M)$. Then
$M_{\la\nu}\ne 0$ iff $\nu\in[\mu]'$, and $M_{\nu\mu}\ne 0$ iff $\nu\in
[\la]$.

\noindent{\it Proof}\quad The proof for (a) was given in \SU\ (it was called
Lemma 3 there); there is no need to repeat the proof here.

{\bf 3(b)} First note that $r=r(B_1)$ (see the proof of Lemma 3 in
\SU). Suppose $r(B_\ell)<r$, for some $\ell$. Look at the sequence of matrices
$({1\over r}M)^j$, for $j\rightarrow\infty$ -- this sequence will not in
general converge. However, from $M'_{\rho\rho}=(S^{(n)\dag}M'S^{(n)}
)_{\rho\rho}$ and $(2.1b)$ we get the very crude bound
$${\rm max}_{\la,\mu}\,M'_{\la\mu}\le \sum_{\la,\mu\in\p^k} M'_{\la\mu}\le
{M'_{\rho\rho}\over S^{(n)2}_{\rho\rho}},\eqno(2.5a)$$
for any positive invariant $M'$. Therefore the entries of $({1\over r}M)^j$,
for any $j$, will be bounded above by $M_{\rho\rho}/(rS^{(n)}_{\rho\rho})$,
so by Bolzano-Weierstrauss the sequence $({1\over r}M)^j$ will have convergent
subsequences. Let $\{M'_i=({1\over r}M)^{j_i}\}_{i=1}^\infty$ denote any
such subsequence, and let $M'$ be its limit.  Clearly, $M'$ will
be a positive invariant (by eqs.(1.4)). The point is that $({1\over r}
B_\ell)^j$ goes to the 0-matrix in the $j\rightarrow \infty$ limit (as can be
seen using Jordan blocks). If $\la$ is one of the weights `in' the block
$B_\ell$, then
$\la\in \p_L(M)$ (because $B_\ell$ is indecomposable and $B_\ell\ne (0)$),
but $\la\not\in\p_L(M')$ (because $({1\over r}B_\ell)^j\rightarrow 0$).
But by hypothesis $M_{\rho\mu}=r\,M'_{\rho\mu}$ for all $\mu$, so by Lemma
1(a) we must have $\p_L(M)=\p_L(M')$ -- a contradiction.

That each $M_{\la\mu}\le r$ follows now by looking at $MM^T$: $r(MM^T)=r(
B_1B_1^T)=r^2$, so $r^2\ge (MM^T)_{\la\la}\ge M_{\la\mu}^2$ by $(2.4c$).

{\bf 3(c)} Let $m=\|\J_L(M)\|$ ($=\|\J_R(M)\|$ by Lemma 1(e)). Write out
the decomposition $\oplus_\ell B_\ell$ for $MM^T$, as in
$(2.4a)$. Let $B_1$ `contain' $\rho$, as usual. Then by hypothesis,
$B_1=B_{(m,m)}$ (see eq.(2.4$d$)), so $r(B_\ell)=m^2$ for each $\ell$ (by
Lemma 3(b)).

Now, let $B_i$ be the block `containing' $\la$. Then for all $J,J'\in
\J_L(M)$, by Lemma 1(c) we get
$$\bigl( B_i\bigr)_{J\la,J'\la}=\bigl(B_i\bigr)_{\la\la}=\sum_{\nu\in\p_R(M)}
M^2_{\la\nu}\ge \sum_{\nu\in[\mu]'}M^2_{\la\nu}=mM^2_{\la\mu}.\eqno(2.5b)$$
Let $B^\la_i$ denote the matrix
$$\bigl(B^\la_i\bigr)_{\nu\nu'}=\left\{\matrix{mM^2_{\la\mu}&{\rm if}\sp
\nu,\nu'\in[\la]\cr 0&{\rm otherwise}\cr}\right. .\eqno(2.5c)$$
Then element-wise, $B_i\ge B_i^\la$ by $(2.5b)$, so by p.57 of \PF{} we
get that $r(B_i)\ge r(B^\la_i)$, with equality iff $B_i=B^\la_i$. But
$B_i^\la=B_{(mM_{\la\mu}^2,m)}$ so $m^2=r(B_i)\ge r(B^\la_i)=m^2M_{\la\mu}^2$.
Therefore $M_{\la\mu}=1$ and $B^\la_i=B_i$.

Using $M^TM$ in place of $MM^T$, we get the corresponding result for $\mu$.
\qquad QED\medskip

Lemma 2 is a corollary of Lemma 3(c).
In Sect.\ 5, Lemma 3(b,c) will be applied to the case $\R^\rho_L(M)=\R^\rho_R
(M)=\O_0\rho$. There is a unique fixed point there: $f=(n/3,n/3)$. So
Lemma 3(c) tells us about most of the weights; the main value of
Lemma 3(b) for us will be in analysing the possible values of $M_{\la f},
M_{f\mu}$, but it will also be useful in Sect.6. Lemma 3 can be strengthened
without difficulty, but this is all we will need in this paper.

The final observation we will use is the {\it parity rule}. Its shortest
derivation is in \CG; there is no need to repeat it here. We will only
state the result, as it applies to SU(3)${}_k$.

For any real numbers $x,y$ define by $\{x\}_y$ the unique number congruent to
$x$ (mod $y$) satisfying $0\le \{x\}_y<y$.
Consider any $\la=(\la_1,\la_2)$, $\la_i\in\Z$, $\la$ not necessarily in
$\p^k$. Define the {\it parity} $\eps(\la)$ of $\la$ to be
$$\eps(\la)=\left\{ \matrix{0 &{\rm if\sp} \{\la_1\}_n,\,\{\la_2\}_n\sp
{\rm or}\sp \{\la_1+\la_2\}_n=0\cr
+1&{\rm if}\sp \{\la_1\}_n+\{\la_2\}_n<n\sp{\rm and}\sp\{\la_1\}_n,\{\la_2\}_n
,\{\la_1+\la_2\}_n>0\cr
-1&{\rm if}\sp \{\la_1\}_n+\{\la_2\}_n>n\sp{\rm and}\sp\{\la_1\}_n,\{\la_2\}_n
,\{\la_1+\la_2\}_n>0\cr}\right. .\eqno(2.6a)$$
Then it can be shown that $\eps(\la)\ne 0$ iff there exists a unique root
lattice vector $v=\ell\,(2,-1)+m\,(1,1)$, $\ell,m\in\Z$, and a unique Weyl
transformation $\omega\in W(A_2)$, such that
$$(\la)^+\equi \omega\la+nv\in\p^k;\eqno(2.6b)$$
in this case $\eps(\la)=det\,\,\omega$.

\medskip\noindent{\bf Lemma 4.} \quad{\bf (a)} Let $M$ be any invariant.
Choose any $\la,\mu\in \p^k$. Then, for each $\ell$ coprime to $3n$, we have
$\eps(\ell\la)\,\eps(\ell\mu)\ne 0$ and
$$M_{\la\mu}=\eps(\ell\la)\,\eps(\ell\mu)\,M_{(\ell\la)^+,(\ell\mu)^+}.
\eqno(2.7)$$

\item{\bf (b)} Now let $M$ be any positive invariant. Then for all
$\ell$ coprime to $3n$, $M_{\la\mu}\ne 0$ implies $\eps(\ell\la)=\eps
(\ell\mu)$.\medskip

The more useful one for our purposes is Lemma 4(b). In fact, we will be
mostly interested in applying it to $\mu=\rho$. This gives us an upper
bound on the sets $\R^\rho_{L,R}(M)$, and from there Lemmas 1,2,3 can be used.
Eq.(2.7) will be used in Sect.6.

The parity rule is extremely powerful. For example,
for the special case $n$ coprime to 6, 4(b) alone is enough to imply
that $\R_L^\la(M)\subset\O\la$, for any positive invariant $M$.
We will not use that here (the
complicated proof is given in \KR). Incidently, a similar result holds
for SU(2)$_k$, $k$ odd -- it is natural to ask how this extends to
higher rank SU($N$).

The `catch' is that proving anything using the parity rule means
immersing oneself in mazes of minute details and special cases. Below is
the parity rule for $\rho$-couplings for the SU(2) modular invariant
classification; it will be used in Sect.4 below, where we will find that
the SU(2) classification is embedded in some way in the SU(3) one.

Let $\C_L$ denote the set of all numbers coprime to $L$.
In order to apply the parity rule, we need a systematic way of producing
lots of numbers $\ell$ in $\C_{L}$. Fortunately, this isn't difficult:
for example, consider
$$\ell={L\over 2^i}\pm 2^j.$$
$\ell$ will lie in $\C_{2L}$ iff either $L/2^i$ is even and $j=0$, or
$L/2^i$ is odd and $j>0$. The reason is that these choices of $i,j$ guarantee
$\ell$ is odd; any other prime $p$ dividing $L$ will not divide $\pm 2^j$
so cannot divide $\ell$.

Choose any integer $m>2$. Let $\R_m$ denote the
set of all integers $a$, $0<a<m$, satisfying:
$$\eqalignno{0<\{\ell\}_{2m}<m\quad{\rm and} &\quad \ell\in\C_{2m}
\sp \Rightarrow\sp\{\ell a\}_{2m}<m;&\cr
m<\{\ell\}_{2m}<2m\quad{\rm and} &\quad \ell \in\C_{2m}
\sp \Rightarrow\sp\{\ell a\}_{2m}>m.&(2.8)\cr}$$

\medskip\noindent{{\bf Lemma 5.}}\footnote{$^1$}{\small This is slightly
more general
than Claim 1 of \SU, since we drop the condition that a must be odd. The proof
is mostly unchanged, apart from many more details included. The proof of
Lemma 5 is fortunately as bad as it gets in this paper!}\quad
 Define the set $\R_m$ as above. Then

\item{(a)} for $m\ne 6$, 10, 12, 30, we have $\R_m=\{1,m-1\}$;

\item{(b)} $\R_6=\{1,3,5\}$;

\item{} $\R_{10}=\{1,3,7,9\}$;

\item{} $\R_{12}=\{1,5,7,11\}$;

\item{} $\R_{30}=\{1,11,19,29\}$.

\noindent{\it Proof}\quad Write $m=2^Lm'$, where $m'$ is odd. Define the
integer $M$ by $m/2\le 2^M<m$. First, note that $a\in\R_m$ iff $m-a\in\R_m$.
The reason is that $\ell\in\C_{2m}$ must be odd, so $\{\ell (m-a)\}_{2m}$
equals $m-\{\ell a\}_{2m}<m$ if $\{\ell a\}_{2m}<m$, or $3m-\{\ell a\}_{2m}>m$
if $\{\ell a\}_{2m}>m$.

Let $a\in\R_m$. Define $b=a/m$, so $0<b<1$, and write out its {\it binary}
(=base 2) {\it expansion}:
$$b=0.b_1b_2b_3\ldots\equi\sum_{i=1}^\infty b_i 2^{-i},\qquad
b_i\in\{0,1\}.\eqno(2.9a)$$
So $b_i=0$ means $\{2^ib\}_2\le 1$, while $b_i=1$ means $\{2^ib\}_2\ge 1$.
For example, ${1\over 2}$ has binary expansion $0.100\ldots=0.011\ldots$, while
${1\over 3}=0.0101\ldots$.

\medskip\noindent{\it Case 1}: $m$ is odd (\ie $L=0$). This is the simplest
case. We may assume
that $a$ is odd (otherwise replace $a$ with $m-a$). Choose $j$ so that
$$2^{-j+1}>b\ge 2^{-j}.\eqno(2.9b)$$
Put $\ell=m-2^j$; it is coprime to $2m$. Then $\{\ell a\}_{2m}=
\{m-2^ja\}_{2m}=3m-2^ja>m$, because $a$ is odd and because of eq.($2.9b)$.
Therefore, by $(2.8$), $\{\ell\}_{2m}>m$, which can only happen if $m-2^j<0$.
Eq.(2.$9b$) now tells us $a<2m/2^j<2$, \ie $a=1$.\medskip

Now, what if $m$ is even? Putting $\ell=m-1$ forces $a\in\R_m$ to be odd: for
if it was even, then $\{\ell a\}_{2m}=2m-a>m$. We may also assume $a\le m/2$,
\ie $b\le {1\over 2}$ (otherwise replace $a$ with $m-a$).

Let $c=\{a/2^L\}_2$. There are two different cases: either $0<c< 1$
(to be called case 2), or $2>c>1$ (to be called case 3). $c\ne 0,1$ because
$a$ is odd and $L>0$. \medskip

\noindent{\it Case 2}: Define $\ell_i=m'+2^i$. Then for $i=1,\ldots,M-1$, each
$\ell_i\in\C_{2m}$, and $0<\ell_i<m$. Then dividing (2.8) by $m$ tells us that
for each $1\le i<M$, either $c+\{2^ib\}_2<1$ or $2<c+\{2^ib\}_2$: more
precisely, for each $i=1,\ldots,M-1$,
$$\{ 2^i b\}_2<1-c\quad{\rm if}\quad \{2^ib\}_2< 1,\qquad{\rm and}\quad
2-c<\{2^i b\}_2\quad{\rm if}\quad \{2^ib\}_2> 1.\eqno(2.10)$$

Choose $j$ so that $(2.9b)$ holds. Suppose
for contradiction that $j<M$. Then by (2.10), $2^{j-1}b=\{2^{j-1}b\}_2<1-c$,
but $2^jb=\{2^jb\}_2> 2-c$. Hence, $2-c< 2^jb<2-2c$, \ie $c<0$, which is false.

Therefore $b<2^{1-M}$, \ie $a<m/2^{M-1}\le 4$, so $a$ odd
implies either $a=1$ or 3. All that remains for case 2 is to show $3\not\in
\R_m$. We will prove this by contradiction. Note that because $c=3/2^L<1$,
we must have $L\ge 2$; $a\le m/2$ then means $m\ge 8$.

If $m\equiv -1$ (mod 3) use $\ell=(m+1)/3$, while if $m\equiv +1$ (mod 3)
use $\ell=(m+2)/3+m'$. If $m\equiv 0,3$ (mod 9) take $\ell=m/3+1$, while
if $m\equiv 6$ (mod 9) use $\ell=m/3+3$. In all cases
$\ell\in\C_{2m}$, $0<\ell<m$, but $\{\ell a\}_{2m}=3\ell>m$, so (2.8) is
violated. Thus $3\not\in\R_m$, and we are done case 2.\medskip

\noindent{\it Case 3}: Take $\ell_i=m'+2^i$ and $\ell_j'=m'-2^j$. Then for
$i,j>0$, both $\ell_i,\ell_j'\in\C_{2m}$. For $i=1,\ldots,M-1$, $0<\ell_i<m$;
for $j=1,\ldots,M-L$, $0<\ell_j'<m$; and for $j=M-L+1,\ldots,M$,
$-m<\ell_j'<0$. Therefore:
$$\eqalignno{{\rm for}\sp 1\le i<M,&\quad 2<\{2^ib\}_2+c<3;&(2.11a)\cr
{\rm for}\sp 1\le j\le M-L,&\quad 0<c-\{2^jb\}_2<1;&(2.11b)\cr
{\rm for}\sp M-L<j\le M,&\quad {\rm either}\sp c<\{2^jb\}_2\sp{\rm or}\sp
1+\{2^jb\}_2<c.&(2.11c)\cr}$$
Now suppose for some $M-L<j<M$, that $c<\{2^jb\}_2$ but $1+\{2^{j+1}b\}_2<c$.
Then $b_j=1$, so $\{2^{j+1}b\}_2=2\{2^jb\}-2$, and we get $c<\{2^jb\}_2<{c+1
\over 2}$, contradicting $c>1$. Similarly, $c<2$ is contradicted if
$1+\{2^jb\}_2<c$ but $c<\{2^{j+1}b\}_2$. Thus
either $c<\{2^ib\}_2$ for all $M-L<i\le M,$ or $1+\{2^ib\}_2<c$ for all
$M-L<i\le M$.
Moreover, if $L>1$ then adding $(2.11a)$ and (2.11$c$) with $i=j=M-1$ tells
us that if $L>1$ then $c<\{2^ib\}_2$ iff $c<{3\over 2}$.

Subtracting $(2.11b)$ from $(2.11a)$ produces the inequality ${1\over 2}<
\{2^ib\}_2<{3\over 2}$ for all $1\le i\le M-L$. Thus, for these $i$, $b_i=0$
iff $b_{i+1}=1$.

Summarizing, we see that the first $M$ binary digits $b_i$ of $b$ are fixed by
the demand that $b_1=0$ (since $a\le m/2$), and eqs.(2.11$a$-$c$):
$$0=b_1\ne b_2\ne b_3\ne\cdots\ne b_{M-L}\ne b_{M-L+1}=\cdots=b_M,
\eqno(2.11d)$$
and for $L>1$, $b_M=1$ iff $c<{3\over 2}$.
This fixes the value of $b$ up to a correction of $2^{-M}$, \ie $a$ up to
$m/2^M\le 2$, so $a$
is then completely fixed by the condition that it be odd. To eliminate
this value of $a$ (except for 7 special values of $m$), we will consider
5 subcases.

(i) Consider first $M-L=1$, \ie $m=3\cdot 2^L$, and $L>2$ (\ie $m\ne 6,12$).
Then by $(2.11d)$ the first $M$ binary digits of $b$ are $b=0.011
\cdots 1$--,
where ``--'' denotes the remaining unknown digits $b_i$, $i>M$. Therefore
$1/2-1/2^M\le b\le 1/2$, \ie
$a=m/2-\eps$ for some $0\le\eps\le m/2^M\le 2$. Therefore $a=m/2-1$. Since
$m>14$, $\ell=7$ lies in $\C_{2m}$ and satisfies $\{\ell
a\}_{2m}={3\over 2}m-7>m$, violating (2.8).

(ii) Consider next $M-L=2$, \ie $m=5\cdot 2^L$ or $7\cdot 2^L$, and $L>2$ (\ie
$m\ne 10,20,14,28$). Then
$b=0.0100\cdots 0$--, \ie $a=m/4+\eps$, for $0\le\eps\le 2$, so $a=m/4+1$.
Then using $\ell=7$ (if $m=5\cdot 2^L$) or $\ell=5$ (if $m=7\cdot 2^L$) will
violate (2.8).

(iii) Now consider $M-L>2$, $L>2$. The first four digits of $b$ will be
$b=0.0101$--. If $c<{3\over 2}$ then putting $j=2$ in $(2.11b)$ gives
$c>1.01$, \ie $c=1.01$--. Now by $(2.11d$), $b_{M-L+1}=\cdots=b_M=1$.
Putting $i=M-L+1$ gives $\{2^ib\}_2+c=1.11$--$+1.01$--$\ge 3$, contradicting
$(2.11a)$.

If instead $c>{3\over 2}$, then putting $j=1$ in $(2.11b$) gives $c=1.10$--.
Eq.$(2.11d$) says $b_{M-L+1}=\cdots=b_M=0$; putting $i=M-L+1$ gives
$\{2^ib\}_2+c=0.00$--$+1.10$--$\le 2$, contradicting $(2.11a$).

(iv) Consider next $M-L>2$, $L=2$. Here, either $c=1.01={5\over 4}$ or
$c=1.11={7\over 4}$. Consider first $c=1.01$, then $b_{M-2}=0$, $b_{M-3}=
b_{M-1}=
b_M=1$ by $(2.11d)$. Putting $j=M-L-1$, we get $c-\{2^jb\}_2=1.01-1.011$--$
<0$, contradicting $(2.11b)$. If instead $c=1.11$, then $j=M-L-1$ gives
$1.11-0.100$--$>1$, also contradicting $(2.11b)$.

(v) Finally(!), look at $M-L>2$, $L=1$. Then $c={3\over 2}$ and $m>16$.
{}From $(2.11d$) we find that for $M$
even ${1\over 3}-{1\over 3\cdot 2^M}\le b\le {1\over 3}+{2\over 3\cdot 2^M}$,
and for $M$ odd ${1\over 3}-{2\over 3\cdot 2^M}\le b\le {1\over 3}+{1\over 3
\cdot 2^M}$. That is, $a=m/3+\eps$, where $-{2\over 3}\le \eps\le{4\over 3}$ if
$M$ is even, and where $-{4\over 3}\le\eps\le{2\over 3}$ if $M$ is odd. So for
$m\equiv 0$ (mod 3), $a_{\pm}=m/3\pm 1$, and if $m\equiv \pm 1$ (mod 3)
we have $a=m/3\mp 1/3$ respectively.

Taking $\ell=3$ in (2.8) eliminates $m\equiv -1$ (mod 3): $\{\ell a\}_{2m}=
m+1>m$.

For $m\equiv +1$ (mod 3), take $\ell=m/2+6$: $\{\ell
a\}_{2m}={10\over 6}m-{1\over 6}m-2={3\over 2}m-2>m$.

All that remains is $m\equiv 0$ (mod 3), \ie $m\equiv 6,18,30$ (mod 36).
For $m\equiv 6$ (mod 36), put $\ell=4+m/6$. $\ell\in
\C_{2m}$, since $\ell\equiv 5$ (mod 6), and $\ell<m$.
$\{\ell a_{\pm}\}_{2m}={4\over 3}m\pm
4+{1\over 3}m\pm {1\over 6}m={11\over 6}m+4$ or ${3\over 2}m-4$, in both
cases violating (2.8).

For $m\equiv 18$ (mod 36), put $\ell=2+m/6$. Again $\ell<m$ and $\ell\equiv 5$
 (mod 6),
so $\ell\in\C_{2m}$. $\{\ell a_{\pm}\}_{2m}={11\over 6}m+2$ or ${3\over 2}m-2$,
so (2.8) is violated.

For $m\equiv 30$ (mod 36), put $\ell=6+m/6$. Again $\ell<m$, and
$\ell\equiv 5$ (mod 6),
so $\ell\in\C_{2m}$. $\{\ell a_{\pm}\}_{2m}={11\over 6}m+6$ or ${3\over 2}m-6$,
so for $m>36$ (\ie $m\ne 30$),  (2.8) is violated.

\medskip
There were some special values of $m$ that slipped through these arguments:
namely $m=6,10,12,14,20,28,30$. These can be worked out explicitly.\qquad QED

\bigskip \bigskip \noindent{{\bf 3. The permutation invariants}}
\bigskip

Recall the definition of permutation invariant given in (2.2$a$). Let
$M^\sigma$ denote its coefficient matrix.
In this section we will find all SU(3)${}_k$ permutation invariants:

\medskip\noindent{{\bf Theorem 1.}} \quad The only level $k$ permutation
invariants for SU(3) are ${\cal A}_k$, ${\cal A}_k^c$ for $k\equiv 0$ (mod 3),
and ${\cal A}_k$, ${\cal A}^c_k$, ${\cal D}_k$, ${\cal D}_k^c$ for
$k\not\equiv 0$ (mod 3). \medskip

That the matrix $M^\sigma$ must commute with $S^{(n)}$
(see (1.4$b$)) is equivalent to
$$S^{(n)}_{\la\mu}=S^{(n)}_{\sigma \la,\sigma \mu},\qquad
\forall \la,\mu\in\p^k.\eqno(3.1)$$
Of course, (P3) tells us $\sigma\rho=\rho$.

Verlinde's formula gives us a relation between the fusion coefficients
$N^{(n)}_{\la\mu\nu}$ and the $S^{(n)}$ matrix:
$$N^{(n)}_{\la\mu\nu}=\sum_{\la'\in \p^k} {S^{(n)}_{\la\la'}S^{(n)}_{\mu\la'}
S^{(n)}_{\nu\la'}\over S^{(n)}_{\rho\la'}}.\eqno(3.2a)$$
We may take $(3.2a)$ as a formal definition of $N^{(n)}_{\la\mu\nu}$.
Eq.(3.1) tells us that
$$N^{(n)}_{\la\mu\nu}=N^{(n)}_{\sigma\la,\sigma\mu,\sigma\nu}.\eqno(3.2b)$$
Eq.(3.2$b$) is useful to us, because these fusion coefficients have been
computed for $SU(3)_k$ \BMW, by a method based on $(3.2a$). The formula is:
$$N^{(n)}_{\la\mu\nu}=\left\{\matrix{0&{\rm if}\sp n\le n_{min} \sp {\rm or}
\sp \delta=0\cr n-n_{min}&{\rm if}\sp n_{min}\le n\le n_{max}\sp {\rm and}
\sp \delta=1 \cr n_{max}-n_{min} & {\rm if} \sp n\ge n_{max}\sp {\rm and}
\sp\delta=1\cr} \right. ,\eqno(3.3)$$
where $\la=(\la_1,\la_2)$, $\mu=(\mu_1,\mu_2)$, $\nu=(\nu_1,\nu_2)$, and
$$\eqalign{A&={1\over 3}[2(\la_1+\mu_1+\nu_1)+\la_2+\mu_2+\nu_2],\cr
 B&={1\over 3}[\la_1+\mu_1+\nu_1+2(\la_2+\mu_2+\nu_2)],\cr
n_{min}&={\rm max}\bigl\{\la_1+\la_2,\mu_1+\mu_2,\nu_1+\nu_2,A-
{\rm min}\{\la_1,\mu_1,\nu_1\},B-{\rm min}\{\la_2,\mu_2,\nu_2\}\bigr\},\cr
n_{max}&={\rm min}\{A,B\},\cr
\delta&=\left\{\matrix{1&{\rm if}\sp n_{max}>n_{min} \sp {\rm and}\sp
 A,B\in \Z\cr 0&{\rm otherwise}\cr} \right. .\cr}$$

The first step in the proof of Thm.1 is to show that `point-wise' $\sigma$
acts like an outer automorphism. To see this, first
take $\la=\mu=\nu=(a,b)$. Then (3.3) reduces to
$$N^{(n)}_{\la\la\la}={\rm min}\{a,b,n-a-b\}.\eqno(3.4a)$$
Eqs.(3.2$b$) and (3.4$a$) imply that
$$\sigma(a,b)\in \bigcup_c\,\{\O(a,c)\cup\O(b,c)\cup \O(n-a-b,c)\}.
\eqno(3.4b)$$
Also, from eq.(3.1) we get $S^{(n)}_{\la\rho}=
S^{(n)}_{\sigma\la,\rho}$. Then the following result forces $\sigma\la
\in\O\la$:

\medskip\noindent{\bf Claim 1.} \quad Choose any $\la=(a,b),\mu=(a',b')\in
\p^k$. If $\{a,b,n-a-b\}\cap\{a',b',n-a'-b'\}\ne\{\}$, and if
$S^{(n)}_{\la\rho}=S^{(n)}_{\mu\rho}$, then $\mu\in\O\la$.

\noindent{\it Proof.} \quad Suppose $(a',b')= A^iC^j(a,c)$ for some
$i,j$. Then $S^{(n)}_{ab,\rho}=
S^{(n)}_{A^iC^j(a,c),\rho}$, \ie $S^{(n)}_{ab,\rho}=S^{(n)}_{ac,\rho}$ by
eqs.(1.6). Eq.(1.3$d$) reduces this to
$$\sin({2\pi b\over n})-\sin({2\pi (a+b)\over n})=\sin({2\pi c\over n})
-\sin({2\pi (a+c)\over n}).\eqno(3.5)$$
Define $f_\alpha(x)=\sin(x)-\sin(x+\alpha)$. We are interested in finding all
solutions $f_\alpha(x)=f_\alpha(y)$, where $x,y,\alpha >0$ and $x+\alpha,y+
\alpha<2\pi$. But since trigonometry tells us $f_\alpha(x)=-2\sin(\alpha/2)\,
\cos(x+\alpha/2)$, it is trivial to solve this. The only solutions in the
given interval are $y=x$ and $y=2\pi-x-\alpha$. Hence the only possible
solutions to (3.5) are: $c=b$ and $c=n-a-b$. In other words,
$$\mu\in \bigcup_c \,\O(a,c)\sp\Rightarrow\sp\mu\in\O(a,b).\eqno(3.6)$$

The calculation and conclusion for $\mu\in\O(b,c)$ is completely
identical to that for $(3.6)$, except that the roles of $a$ and $b$ are
interchanged. Similarly for $\mu\in\O(n-a-b,c)$, except with $a$
above replaced with $n-a-b$ here. \qquad QED to Claim 1\medskip

The claim, together with (2.1$c$), tells us that the only possibilities for
$\sigma(1,2)$ are:
$$\eqalignno{\sigma(1,2)\in &\bigl\{(1,2),(2,1)\bigr\}\sp {\rm if}\sp k\equiv
0\sp{\rm (mod}\sp 3),&(3.7a)\cr
\sigma(1,2)\in &\bigl\{(1,2),(2,1),(2,k),(k,2)\bigr\}\sp {\rm if}\sp k\equiv
1\sp{\rm (mod}\sp 3),&(3.7b)\cr
\sigma(1,2)\in &\bigl\{(1,2),(2,1),(k,1),(1,k)\bigr\}\sp {\rm if}\sp k\equiv
2\sp{\rm (mod}\sp 3).&(3.7c)\cr}$$
Note that the possibilities for $k\equiv 0$ (mod 3) are realized by
${\cal A}_k$ and ${\cal A}_k^c$, respectively, and for $k\equiv \pm 1$ (mod 3)
by ${\cal A}_k$, ${\cal A}_k^c$, ${\cal D}_k$ and ${\cal D}_k^c$, respectively.
Since the (matrix) product $M^\sigma M^{\sigma'}$ of two permutation invariants
$M^\sigma$, $M^{\sigma'}$ is another permutation invariant $M^{\sigma'
\sigma}$, to prove Thm.1 for each $k$ it suffices to show that the only
permutation invariant satisfying $\sigma(1,2)=(1,2)$ is ${\cal A}_k$.

Choose any $\la\in\p^k$. Then $\sigma\la=C^iA^j\la$ for some $i,j$, by Claim 1.
Then $(3.1)$ tells us $S^{(n)}_{(1,2),\la}=S^{(n)}_{(1,2),C^iA^j\la}$. By the
following claim this forces $\sigma\la=C^iA^j\la=\la$.\qquad QED to Thm.1

\medskip\noindent{\bf Claim 2.}\quad For any $\la\in\p^k$, $S^{(n)}_{(1,2),\la}
=S^{(n)}_{(1,2),C^iA^j\la}$ iff $\la=C^iA^j\la$.

\noindent{\it Proof}\quad Suppose for some $1\le a<n-1$ that  $S^{(n)}_{12,1a}
=S^{(n)*}_{12,1a}$. Then
$$c_n(5a+4)+c_n(a+5)+c_n(4a-1)=c_n(4a+5)+c_n(5a+1)+c_n(a-4),\eqno(3.8a)$$
where $c_n(x)=\cos(2\pi {x\over 3n})$. Making the substitution $b=a+
{1\over 2}$, we would like to show that
$$p(b,n)\equi c_n(5b+{3\over 2})+c_n(b+{9\over 2})+c_n(4b-3)-c_n(4b+3)-
c_n(5b-{3\over 2})-c_n(b-{9\over 2})\eqno(3.8b)$$
does not vanish for ${3\over 2}<b<n-{1\over 2}$.

Using the obvious trigonometric identities (\ie the angle-sum formulas, and
sin$^2=1-$cos$^2$) we can rewrite $p(b,n)$ as a
polynomial in $c_n(b)$ and $s_n(b)=\sin(2\pi {b\over 3n})$ -- in particular,
$p(b,n)=p^{(n)}_5\bigl(c_n(b)\bigr)+s_n(b)\cdot p^{(n)}_4\bigl(c_n(b)\bigr)$,
where $p^{(n)}_5$ and $p^{(n)}_4$ are, respectively, degree 5 and 4
polynomials. Note from (3.8$b$) that $p(-b,n)=-p(b,n)$, so $p^{(n)}_5$ must
be identically zero:
$$p(b,n)=s_n(b)\cdot p^{(n)}_4\bigl(c_n(b)\bigr).\eqno(3.8c)$$
We are interested in the roots of this function, in the range $b\in(0,{3\over
2}n)$. Since for these $b$ $s_n(b)$ does not vanish and $c_n$ is one-to-one,
for fixed $k$ there can be at most 4 zeros for $p^{(n)}_4$ and hence
$p(b,n)$ in that range. But $b={1\over 2},{3\over 2},n-{1\over 2},n+{1\over 2}$
are 4 distinct zeros for $p(b,n)$. Therefore they are the {\it only} zeros
in the range $b\in (0,{3\over 2}n)$, and so $p(b,n)$ cannot vanish when
${3\over 2}<b<n-{1\over 2}$.

This argument is quite general. It isn't necessary to assume $a$ is an
integer; $(1.3d)$ can be used to formally define $S^{(n)}_{\la\mu}$ for
arbitrary $n,\la,\mu$. All that is required for our argument is that
$n>1$ (so that the root $b=n+{1\over 2}$ satisfies $0<b<{3\over 2}n$) and
$n\ne 2$ (so $n-{1\over 2}\ne {3\over 2}$).

Now choose any $\la\in\p^k$. Wlog suppose $\la_1\le \la_2$. Put $n'=
n/\la_1$ and $a=\la_2/\la_1$. Then $1\le a<n'-1$ and $n'>2$. We see
that $S^{(n)}_{(1,2),\la}={n'\over n}S^{(n')}_{(1,2),(1,a)}$.
This means we have the equation $S^{(n')}_{(1,2),(1,a)}=S^{(n')*}_{(1,2),
(1,a)}$. But from the first part of this argument, we know that this
is possible only for $a=1$, \ie $\la_1=\la_2$.

Now suppose $S^{(n)}_{(1,2),\la}=0$. Then by (1.6$d$),
$S^{(n)}_{(1,2),A^i\la}=0$ for all $i$. Because these are all real,
$(A^i\la)_1=(A^i\la)_2$ for all $i$, \ie $\la_1=\la_2=n-\la_1-\la_2$,
\ie $\la=(n/3,n/3)$.

Finally, suppose $S^{(n)}_{(1,2),C^iA^j\la}=S^{(n)}_{(1,2),\la}$.
Consider first the case where $i=1$. We get
$$S^{(n)}_{(1,2),\la}=\omega^j S^{(n)*}_{(1,2),\la},\qquad i.e.\quad
S^{(n)}_{(1,2),A^{-j}\la}=S^{(n)*}_{(1,2),A^{-j}\la},\eqno(3.9a)$$
where $\omega=\exp[2\pi i/3]$. We know this implies $(A^{-j}\la)_1
=(A^{-j}\la)_2$, \ie $CA^{-j}\la=A^{-j}\la$. Therefore $\la=A^jCA^{-j}\la=
CA^{-j}A^{-j}\la=C^iA^j\la$.

Now suppose $i=0$. Then
$$S^{(n)}_{(1,2),\la}=\omega^{-j}S^{(n)}_{(1,2),\la}.\eqno(3.9b)$$
If $j\equiv 0$ (mod 3), then trivially $C^iA^j\la=\la$. Otherwise, $(3.9b)$
requires $S^{(n)}_{(1,2),\la}=0$. We know this means $\la=(n/3,n/3)$,
so again $C^iA^j\la=\la$.\qquad QED to Claim 2

\bigskip \bigskip \noindent{{\bf 4. The $\rho$-couplings}}
\bigskip

Define
$${\cal R}^k=\bigcup_M \,\R^\rho_L(M)=\bigcup_M\,\R^\rho_R(M),$$
the unions being over all positive invariants $M$ of $SU(3)_k$.
For example, the known SU(3) physical invariants (1.7) tell us that
${\cal R}^5\supseteq \{(1,1),(3,3)\}$, ${\cal R}^6\supseteq \{(1,1),(7,1),
(1,7)\}$ and ${\cal R}^7\supseteq\{(1,1)\}$. ${\cal R}^k$ is called the set of
$\rho$-{\it couplings} at level $k$. We learned in Sect.2 that the
$\rho$-couplings should be both accessible and informative. For instance,
if ${\cal R}^k=\{\rho\}$ then by Lemma 2 any level $k$ physical invariant
will be a permutation invariant, and will be listed in Thm.1.

Let $\la=(a,b)\in {\cal R}^k$. Then by $(2.1c$) it must satisfy
$$a^2+ab+b^2 \equiv  3 \sp ({\rm mod\sp} 3n).\eqno(4.1a)$$
Another important property comes from Lemma 4(b):
$$\eqalignno{0<\{\ell\} <n/2\quad{\rm and}\quad&\ell\in\C_n\quad
 \Rightarrow \quad\{\ell a\}+\{\ell b\}<n,&\cr
n/2<\{\ell\} <n\quad{\rm and}&\ell\in\C_n\quad\Rightarrow\quad
 \{\ell a\}+\{\ell b\}>n.&(4.1b)\cr}$$
Two comments about $(4.1b$) must be made. One is that, throughout this
section, we will write $\{\cdots\}$ for $\{\cdots\}_n$. The other is that we
write in $(4.1b$) that $\ell\in\C_n$, not $\ell\in\C_{3n}$. The reason is that
for any $\ell\in\C_n$, there can be found an $\ell'\in\C_{3n}$ such that
$\ell\equiv\ell'$ (mod $n$); from $(2.6a)$ we see that $\eps(\ell\la)=
\eps(\ell'\la)$ for any $\la$.

This section is devoted to a proof of:

\medskip\noindent{\bf Theorem 2.}\quad The only solutions to eqs.(4.1) are:

\item{(i)} for $k\equiv 2,4,7,8,10,11$ (mod 12):
$$(a,b)\in\{(1,1)\};\eqno(4.2a)$$

\item{(ii)} for $k\equiv 1,5$ (mod 12):
$$(a,b)\in \{(1,1),\,({k+1\over 2},{k+1\over 2})\};\eqno(4.2b)$$

\item{(iii)} for $k\equiv 0,3,6$ (mod 12):
$$(a,b)\in \{(1,1),\,(1,k+1),\,(k+1,1)\};\eqno(4.2c)$$

\item{(iv)} for $k\equiv 9$ (mod 12), $k\ne 21,57$:
$$(a,b)\in \{(1,1),\,(1,k+1),\,(2,{k+1\over 2}),\,(k+1,1),\,({k+1\over 2},2),
\,({k+1\over 2},{k+1\over 2})\};\eqno(4.2d)$$

\item{(v)} for $k=21$ and $k=57$, resp.:
$$\eqalignno{(a,b)\in&\,\O_0(1,1)\,\cup\,\O_0(5,5)\,\cup\,\O_0(7,7)\,\cup\,\O_0
(11,11);&(4.2e)\cr
(a,b)\in&\,\O_0(1,1)\,\cup\,\O_0(11,11)\,\cup\,\O_0(19,19)\,\cup\,\O_0(29,29).
&(4.2f)\cr}$$

The reason $k=21$ and $k=57$ are singled out here turns out to be the same
(see Lemma 5) as the reason $k=10$ and $k=28$ are singled out in the
corresponding $\rho$-couplings for $SU(2)_k$. Indeed, $21+3=2(10+2)$ and
$57+3=2(28+2)$. This embedding of the $SU(2)_k$ classification inside
the $SU(3)_{2k+1}$ one remains a mystery, at least to me!

We will prove Thm.2 later in the section. For now let us consider what
would happen if it were true. It gives an upper bound for the sets ${\cal
R}^k$.
So for half of the levels, Thm.2 reduces the completeness proof to the
classification of the permutation invariants, which was accomplished in
Thm.1. Thm.2 turns out to be
sufficient to complete the $SU(3)_k$ classification for all $k$ (this is
done in Sect.5).

\medskip
\noindent{{\bf Claim 3.}} For any $k$ and any $\la\in \p^k$, $\la$
satisfies the parity condition (4.1$b$) iff every $\la'\in\O\la$ does.
Moreover,  if $\la=(a,b)$ satisfies the condition
$$a^2+ab+b^2\equiv 3 \sp({\rm mod}\sp n),\eqno(4.3)$$
then so will every $\la'\in\O\la$.\medskip

The proof of Claim 3 is a straightforward calculation. For example,
 if $\{\ell a\}+\{\ell
b\}<n$, then $\{\ell n-\ell a-\ell b\}+\{\ell a\}=n-\{\ell a\}-\{\ell b\}
+\{\ell a\}=n-\{\ell b\}<n$; while if $\{\ell a\}+\{\ell b\}>n$ then the
same calculation gives $\{\ell n-\ell a -\ell b\}+\{\ell a\}=2n-\{\ell b\}>n$.

Because of Claim 3, we will restrict our attention for the remainder of this
section to any weight $(a,b)\in \p^k$ satisfying the parity condition $(4.1b)$
and the norm condition (4.3). By Claim 3 this set of possible $\la$ is
invariant under the outer automorphisms $\O$. What we will actually prove is
the simpler (and more general):

\medskip \noindent{\bf Proposition 1.} The set of all solutions $\la\in\p^k$ to
$(4.1b)$ and $(4.3)$, where $n=k+3$, is:

\item{(a)} for $n\equiv 1,2,3$ (mod 4), $n\ne 18$: \quad $\la\in\O_0\rho$;

\item{(b)} for $n\equiv 0$ (mod 4), $n\ne 12, 24,60$: \qquad $\la\in\O_0\rho\,
\cup\,\O_0(\rho')$, where $\rho'=({n-2\over 2},{n-2\over 2})$;

\item{(c)} for $n=12,18,24,60$, respectively, $\la$ lies in
$$\eqalign{\O_0\rho&\,\cup\,\O_0(3,3)\,\cup\,\O_0(5,5),\cr
\O_0\rho&\,\cup\,\O(1,4),\cr
\O_0\rho&\,\cup\,\O_0(5,5)\,\cup\,\O_0(7,7)\,\cup\,\O_0(11,11),\cr
\O_0\rho&\,\cup\,\O_0(11,11)\,\cup\,\O_0(19,19)\,\cup\,\O_0(29,29).\cr}$$

\noindent{{\it Proof of Prop.1 when $n\equiv 0$.}} \quad We learn
from
the norm condition ($4.3$) that two of $a$, $b$ and $n-a-b$ will be odd and
one will be even; from Claim 3 we may assume for now that both $a$ and $b$ are
odd. Let $0<\ell<n/2$, $\ell\in\C_n$. Then $\ell'=\ell
+n/2$ will also lie in $\C_n$ but will fall in the range
$n/2$ to  $n$. Then ($4.1b$) tells us
$$\{\ell a\}+\{\ell b\}<n<\{\ell' a\}+\{\ell' b\}.\eqno(4.4)$$
But $a$ is odd, so $\{\ell' a\}=\{n/2+\ell a\}$ equals
$n/2+\{\ell a\}$ if $\{ \ell a\}<n/2$, or $-n/2+\{\ell a\}$ if $\{\ell a
\}>n/2$. A similar comment applies to $b$. If $n/2<\{\ell a\}$, then
$\{\ell'a\}+\{\ell'b\}= -n/2+\{\ell a\}\pm n/2+\{\ell b\}\le \{\ell a\}
+\{\ell b\}$, contradicting (4.4); similarly with $b$. So $\{\ell a\},\{\ell
b\}<n/2$. Thus, putting $m=n/2$ we get exactly the situation stated in
Lemma 5. From there we read that the only possibilities for $a$ and $b$ are
1 and $(n-2)/2$, unless $n=12,20,24,60$. From these we can also compute the
possibilities for $n-a-b$. Eq.(4.3) now reduces this list of
possibilities to those given in the Proposition. \qquad {\bf QED} when 4
divides $n$\bigskip

Thus it suffices to consider $n\equiv 1,2,3$ (mod 4). First we
will prove two useful results.

\medskip\noindent{{\bf Claim 4.}} For $n\equiv 1,2,3$ (mod 4), if $a=b$
then $a=b=1$.

\noindent{{\it Proof.}} \quad  Clearly $a<n/2$, since $a+b<n$.
For $n$ even, $(4.1b$) reduces to the hypothesis of Lemma 5 with $m=n/2$,
and we  get $a=1$.

Otherwise, $n$ is odd.
Let $N>0$ be the unique integer for which $2^N<n/2<2^{N+1}$. Similarly,
let $j\ge 0$ be the smallest integer for which $2^ja<n/2<2^{j+1}a$.
Assume for contradiction that $a>1$. Then $0\le j<N$. Take $\ell=2^{j+1}<n/2$.
Then we get $\{\ell a\}+\{\ell b\}=2(2^{j+1}a)>n$, contradicting
(4.1$b$). \qquad QED to Claim 4\medskip

\noindent{{\bf Claim 5.\footnote{$^2$}{\small The proof of this claim given
in \SU{} was cryptic enough to have given a lot of people problems. The idea
I had in mind was quite geometrical, and essentially along the lines given
here. The statement of the claim here is actually slightly more general,
since it uses only (4.3) -- this is also the reason for the n=12 exception
here.}}}\quad The greatest common divisors $gcd(a,n)$, $gcd(b,n)$,
$gcd(n-a-b,n)=gcd(a+b,n)$, equal either 1 or 2 (except for $n=12)$.

\noindent{{\it Proof.}} \quad By Claim 3 it suffices to prove $gcd(a,n)\le 2$.
Suppose for contradiction a prime $p\ne 2$ divides both $a$ and $n$. Consider
first $p=3$. Then (4.3) tells us that 3 must also divide $b$, and that 9
cannot divide $n$ (otherwise (4.3) would imply $0\equiv 3$ (mod 9)). Look at
$\ell_m=3+mn/ 3$; then $\ell_1,\ell_2\in\C_n$,
and for $n>18$, $0<\ell_1<n/2<\ell_2<n$. However, $\{\ell_ma\}+\{\ell_mb\}
=\{3a+mn{a\over 3}\}+\{3b+mn{b\over 3}\}=\{3a\}+\{3b\}$ is independent of $m$,
so either $\ell_1$ or $\ell_2$ must violate $(4.1b$). (The remaining heights
$n=6,12,15$ can be checked by hand; only $n=12$ turns out to allow 3 to divide
$a$.)

Now consider other primes $p$. By (4.3), $b^2\equiv 3$ (mod $p$), which has
no solutions if $p\le 7$.
Let $\ell_m=1+mn/p$, $m=0,1,\ldots,p-1$. Then $\ell_m\in\C_n$ iff $\ell_m
\not\equiv 0$ (mod $p$), so except possibly for one value of
$m$, call it $m_0$, each $\ell_m$ will lie in $\C_n$. Since $p$ divides both
$a$ and $n$, and $0<a<n$, we know $n\ne p$. Therefore $0<\ell_m<n/2$ for all
$0\le m\le {p-1\over 2}$, and $n/2<\ell_m<n$ for ${p+1\over 2}\le m\le p-1$.
So from $(4.1b$) we get
$$\{b+m{bn\over p}\}<n-a\eqno(4.5a)$$
for each $m$ satisfying $0\le m\le{p-1\over 2}$ (except possibly $m=m_0$), and
$$\{b+m{bn\over p}\}\ge n-a\eqno(4.5b)$$
for each $m$ satisfying ${p+1\over 2}\le m\le p-1$ (except possibly $m=m_0$).
An `=' has been added to $(4.5b)$, purely for later convenience.

Now define $b'=\{b+{p\over 2}\}_p-{p\over 2}$ (so $b'\equiv b$ (mod $p$) and
$-{p\over 2}\le b'<{p\over 2}$). The geometric picture: each time $m$ is
incremented by 1, $\{\ell_mb\}=\{b+m{b'n\over p}\}$ changes by precisely
${b'n\over
p}$, unless $\ell_mb$ ``crosses'' an integer multiple of $n$, in which case
$\{\ell_mb\}$ changes by $\pm n+{b'n\over p}$.
{}From this picture we can see that if each $m$ in $m_1\le m\le m_2$ satisfies
$(4.5a)$, then
$$|b'|{n\over p}(m_2-m_1)<n-a, \eqno(4.5c)$$
unless $|b'|{n\over p}>a$;
similarly, if each $m$ in $m_3\le m\le m_4$ satisfies $(4.5b)$, then
$$|b'|{n\over p}(m_4-m_3)<a,\eqno(4.5d)$$
unless $|b'|{n\over p}>n-a$.
We will use $(4.5c,d)$ to prove $|b'|\le 1$; this would mean $b\equiv +1,0,-1$
(mod $p$), none of which can satisfy (4.3) (which tells us $b^2\equiv 3$ (mod
$p$)).

First let us eliminate the possibilities $|b'|{n\over p}>a$ and $|b'|{n\over p}
>n-a$. The former would violate $(4.5d$) if we had an $m$ such that both
$m$ and $m+1$ satisfied $(4.5b$) (since if also $|b'|{n\over p}>n-a$, this
would contradict $|b'|\le {p\over 2}$). But there are ${p-1\over 2}$ numbers
between ${p+1\over 2}$ and $p-1$ inclusive, only one of which can equal $m_0$.
Therefore if ${p-1\over 2}>3$, \ie $p>7$, such an $m$ will exist. Similarly,
to eliminate $|b'|{n\over p}>n-a$ requires that $p>5$.

Suppose now that $0<m_0<{p-1\over 2}$, and that $m_0$ satisfies $(4.5b)$.
Then $\{\ell_{m_0\pm 1}b\}<n-a\le \{\ell_{m_0}b\}$, which forces $2|b'|{n\over
p}>a$ by our geometric picture. But because $p>7$, ${p+5\over 2}\le p-1$,
so $m={p+1\over 2},{p+3\over 2},{p+5\over 2}$ all must satisfy $(4.5b)$.
Eq.(4.5$d$) now tells us $2|b'|{n\over p}<a$, and we get a contradiction.
Thus if
$0<m_0<{p-1\over 2}$, then $m_0$ satisfies $(4.5a$). The identical argument
shows that if ${p+1\over 2}<m_0<p-1$, then $m_0$ must satisfy $(4.5b)$.

Thus there are exactly 4 possibilities ($m_0=0$ cannot happen):

\item{(i)} eq.(4.5$a$) holds for all $0\le m\le{p-1\over 2}$, and $(4.5b)$
holds for all ${p+1\over 2}\le m\le p-1$;

\item{(ii)} $m_0={p-1\over 2}$, and $m_0$ satisfies $(4.5b)$;

\item{(iii)} $m_0={p+1\over 2}$, and $m_0$ satisfies $(4.5a)$;

\item{(iv)} $m_0=p-1$ and $m_0$ satisfies $(4.5a)$.

In case (i) define $m_1=0$, $m_2={p-1\over 2}$; in (ii) define $m_1=0$, $m_2
={p-3\over 2}$; in (iii) define $m_1=0$, $m_2={p+1\over 2}$; and in (iv)
define $m_1=-1$, $m_2={p-1\over 2}$. Then in all four cases we have ($4.5a$)
satisfied for all $m_1\le m \le m_2$, and ($4.5b)$ satisfied for all
$m_2<m\le p+m_1-1$. Then (4.5$c,d$) tell us $|b'|{n\over p}(m_2-m_1)<n-a$
and $|b'|{n\over p}(p+m_1-1-m_2-1)<a$. Adding these gives $|b'|{n\over p}
(p-2)<n$, \ie $|b'|<{p\over p-2}<2$, \ie $|b'|\le 1$. But as we said this
contradicts (4.3).

Therefore, $p=2$ is the only prime that can divide both $a$ and $n$. Since
$(4.3$) shows 4 cannot divide both (since then $b^2\equiv 3$ (mod 4), which
has no integer solutions), the only possibilities for $gcd(a,n)$
are 1 or 2. \qquad QED to Claim 5

\bigskip \noindent{{\it Proof of Prop.1 for $n$ odd.}}\quad
{}From Claim 3 we may assume $1\le a,b< n/2$. We need to show $a=b=1$.

First take $\ell=(n-1)/2$; it lies in $\C_n$ and is less than $n/2$.
If $a$ is even, $\{\ell a\}=n-a/2$, and if $a$ is odd, $\{\ell a\}=n/2
-a/2$. The same applies for $b$. Hence $\{\ell a\}+\{\ell b\}=in+(n-a-b)/2$,
where $i=1/2,1,3/2$
depending on whether 0, 1 or both of $a,b$ are even. But $i\ge 1$ contradicts
(4.1$b$) -- since as always $a+b<n$. Therefore both $a$ and $b$ must be odd.

Eq.(4.3) tells us $\{a^2\}+\{ab\}+\{b^2\}=3+mn$, for some integer $m$.
Since by definition $0\le \{\cdots\}<n$, we have $m=0,1,$ or 2. But
$m=2$ would imply $\{a^2\}+\{ab\}=3+2n-\{b^2\}>n$,
which contradicts (4.1$b$) with $\ell=a$ ($a<n/2$ by hypothesis, and $a\in\C_n$
by Claim 5).

Next suppose $m=1$, \ie
$$\{a^2\}+\{ab\}+\{b^2\}=n+3.\eqno(4.6)$$
Choose $\ell'=
(n+a)/2$, $\ell''=(n+b)/2$ -- again Claim 5 tells us these lie in $\C_n$, and
both satisfy $n/2<\ell',\ell''<n$. Then $\ell' a\equiv n/2+a^2/2$ (mod $n$),
so $\{\ell' a\}=\{a^2\}/2+n/2$ if $\{a^2\}$ is odd, and $\{a^2\}/2$ if
$\{a^2\}$ is even. Similarly, $\{\ell' b\}=\{\ell''a\}=\{ab\}/2+n/2$ or
$\{ab\}/2$, depending on whether $\{ab\}$ is odd or even, resp., and
$\{\ell''b\}=\{b^2\}/2+n/2$ if $\{b^2\}$ is odd, and $\{b^2\}/2$ if $\{
b^2\}$ is even. But (4.6) tells us that $\{a^2\}+\{ab\}+\{b^2\}$ is even,
so either all three are even, or 2 are odd and 1 is even. If $\{a^2\}$ or
$\{ab\}$ (or both) are even, then using $\ell'$ in (4.1$b$) gives $n<\{\ell'
a\}+\{\ell' b\}\le\{a^2\}/2+n/2+\{ab\}/2$, \ie $n<\{a^2\}+\{ab\}$, but
this contradicts $(4.1b$) with $\ell=a$ chosen (by hypothesis $a<n/2$, and
$a\in\C_n$ by Claim 5). Similarly, if instead $\{b^2\}$ is even, then using
$\ell''$ in $(4.1b$) contradicts using $b$ in $(4.1b$).

Thus $m=0$ is forced. This requires $\{a^2\}=\{ab\}=\{b^2\}=1$, \ie
$a^2\equiv ab\equiv b^2\equiv 1$ (mod $n)$; Claim 5 tells us $a$ is invertible
(mod $n$), so $a^2\equiv ab$ (mod $n$) implies $a\equiv b$ (mod $n$), \ie
$a=b$. Claim 4 now forces $a=b=1$.
\qquad {\bf QED} to Prop.1 for $n$ odd \bigskip

\noindent{{\it Proof of Prop.1 for $n\equiv 2$
(mod 4).}}\footnote{$^3$}{\small This argument was
quite complicated in \SU; it has been completely rewritten here. I think this
proof is more natural, but probably can still be simplified. The basic idea
is simple: we make 4 series of numbers coprime to n out of powers of 2;
writing down the (4.1b) inequalities for these forces either a=b=1 or n=18.
It is the intricate and not very interesting details which make this argument
so long.}\quad This is the final possibility.
{}From (4.3) we get that both $a$ and $b$
cannot be even, so by Claim 3 we may assume $a,b$ are both odd, and that
$a\le b$. Then $a+b<n$ implies $a<n/2$, so $\{a^2\}+\{ab\}<n$ by $(4.1b$).
We want to show $a=b=1$.

Now, exactly as in the proof for $n$ odd, $\{a^2\}+\{ab\}+\{b^2\}=2n+3$
contradicts $(4.1b$) with $\ell=a$ chosen. Also, $\{a^2\}+\{ab\}+\{
b^2\}=3$ requires $a=b$ and hence $a=b=1$, again exactly as in
the proof for $n$ odd.

Thus it suffices to consider the case where (4.6) is satisfied.
Define $M$ by $2^M<n/2<2^{M+1}$, so $n/2^M<4$. As in $(2.9a$),
write out the binary expansions
$a/n=\sum_{i=1}^\infty a_i 2^{-i}$, $b/n=\sum_{i=1}^\infty b_i 2^{-i}$,
where each $a_i,b_i\in \{0,1\}$.  Note that we cannot have all but
finitely many $a_i$ or $b_i$ equal to 1, say (same for 0), because that
would mean $a/n$ or $b/n$, respectively, was a dyadic rational (\ie its
denominator is a power of 2) -- but $n\equiv 2$ (mod 4), so this would force
$a=n/2$ or $b=n/2$, which contradicts Claim 5.

Consider $\ell_i=n/2+2^i$, $i=1,\ldots,M$. Then $\ell_i\in\C_n$, and
$n/2<\ell_i<n$, so by $(4.1b)$
$$n<\{\ell_i a\}+\{\ell_i b\}=\{{n\over 2}+2^ia\}+\{{n\over 2}+2^ib\}=
\{2^ia\}+\{2^ib\}+\left\{\matrix{n&{\rm if}\sp a_{i+1}=b_{i+1}=0\cr
0&{\rm if}\sp a_{i+1}+b_{i+1}=1\cr -n&{\rm if}\sp a_{i+1}=b_{i+1}=1\cr}
\right. \eqno(4.7a)$$
The reason for $(4.7a$) is that $\{2^ia\}>n/2$ iff $a_{i+1}=1$ (similarly
for $b$). Now, $\{\cdots\}<n$, so $(4.7a)$ forbids $a_{i+1}=b_{i+1}=1$, for all
$i=1,2,\ldots,M$ (the relation $a+b<n$ forbids it for $i=0$).

Define $I$ by $n/2^I<b<n/2^{I-1}$, \ie $b_i=0$ for $i<I$ and $b_I=1$.
If $I-1>M$, then $a\le b<2$, \ie $a=b=1$. So we may suppose $I-1\le M$.

Consider first the case $I>1$. Then ($4.7a$) with $i=I-1$ tells us
$n<\{2^{I-1}a\}+\{2^{I-1}b\}=2^{I-1}a+2^{I-1}b$, \ie $n/2^{I-1}<a+b$. This
is a strong inequality because the biggest $a+b$ can be is if $a_i+b_i=1$ for
$I\le i\le M+1$, and $a_j=b_j=1$ for $j>M+1$: this leads to the bound
$a+b<n/2^{I-1}+n/2^{M+1}$. But if instead $a_i=b_i=0$ for some $i\le M+1$,
then $a+b< n/2^{I-1}+n/2^{M+1}-n/2^i\le n/2^{I-1}$, contradicting the
$\ell_{I-1}$ result. Thus $a_i+b_i=1$ is indeed forced for all $i\le M+1$:
\ie
$$I>1\Rightarrow a+b={n\over 2^{I-1}}+\epsilon,\sp {\rm where}\sp 0<
\epsilon<n/2^{M+1}<2.\eqno(4.7b)$$

The case $I=1$ is identical ($\{2^ib\}$ is independent of $b_1$, for $i\ge
1$). Define $I'>1$ to be the smallest index (other
than $I=1$) with $a_{I'}=1$ or $b_{I'}=1$. Again $I'-1\le M$, because otherwise
$n/2<b<n/2+2$, impossible since $b$ is odd. Then the identical argument gives
$$I=1\Rightarrow a+b={n\over 2}+{n\over 2^{I'-1}}+\epsilon,\sp {\rm where}
\sp 0<\epsilon<2\eqno(4.7c)$$
In both $(4.7b,c$), $\epsilon$
is fixed by the constraint that $a+b$ must be even. Thus we have essentially
removed one degree of freedom. First we will constrain $I,I'$.

\medskip \noindent{\bf Claim 6.}\quad $I\ne 2$. If $I=1$ then $3\le I'<M$
(unless $n=18$), and
$$\{ab\}+\{b^2\}=n+2,\qquad \{a^2\}=1.\eqno(4.8)$$

\noindent{\it Proof.}\quad Suppose first that $I=2$. Then $a+b=n/2+1$, by
$(4.7b$). From this, we can compute $a^2$, $ab$, $b^2$ (mod $n$): $ab\equiv
(a+b)^2-a^2-ab-b^2\equiv n/2-2$; $a^2\equiv (a+b)a-ab\equiv a+2$;
$b^2\equiv (a+b)b-ab\equiv b+2$. $a^2\equiv a+2$ tells us
either ${a(a-1)\over 2}\equiv 1$ (mod $n$) (if $a\equiv -1$ (mod 4)), or
${a(a-1)\over 2}+{1\over 2}n\equiv 1$ (mod $n$) (if $a\equiv +1$ (mod 4)).
Then $a\equiv +1$ (mod 4) would violate $(4.1b$) with $\ell={a-1\over 2}+
{n\over 2}$ (for then $\{\ell a\}=1$; $\ell\in\C_n$ because we have learned
${a-1\over 2}\in \C_{n/2}$), so
$a\equiv -1$ (mod 4). Similarly, we must have $b\equiv -1$ (mod 4), so
${n\over 2}+1=a+b\equiv 2$ (mod 4), \ie $n\equiv 2$ (mod 8). Now take $\ell=
{n+2\over 4}$; $a,b<n/2$ so we get $\{\ell a\}+\{\ell b\}=\{{-n\over 4}+
{a\over 2}\}+\{{-n\over 4}+{b\over 2}
\}={3n\over 2}+{n+2\over 4}>n$, contradicting $(4.1b$).

Now suppose $I=1$. Then by $(4.7c$), $I'=2$ would violate $a+b<n$. $I=1$
means $b>n/2$, so $(4.1b$) gives $\{ab\}+\{b^2\}>n$. But by (4.6), and the
fact that both $\{ab\}$ and $\{b^2\}$ must be odd, we get (4.8).

If $I'\ge
M$, then $a<n/2^{I'-1}\le n/2^{M-1}<8$, and $n/2<b<n/2+8$, which give us the
possibilities $a=1$, 3, 5, or 7, and $b=n/2+2$, $n/2+4$, or $n/2+6$. But
for $a=3,5,7$ respectively, the condition $\{a^2\}=1$ means $n$ must
divide 8, 24, and 48.  $(a,b)=(1,n/2+2)$ means $2\equiv ab+b^2\equiv
n/2+2+n/2+4$ (mod $n$), \ie $n$ divides 4. $(a,b)=(1,n/2+4)$ means $2\equiv
n/2+4+n/2+16$ (mod $n$), \ie $n$ divides 18. $(a,b)=(1,n/2+6)$ means $2\equiv
n/2+6+n/2+36$ (mod $n$), \ie $n$ divides 40.

But $n>3$ and $n\equiv 2$ (mod 4), so the $n$ which must be explicitly
checked are $n=6,10,18$. \qquad QED to Claim 6\medskip

Now that we know so much about $a$, $b$ and $a+b$, a similar game can
be played with them. In particular, define $\ell_i'=\{n/2+2^ia\}$, $\ell''_i
=\{n/2+2^ib\}$, and $\ell'''_j=\{n/2+2^j(a+b)\}$. Note that from Claim 5 these
lie in $\C_n$ for $i\ge 1$ and $j\ge 0$.

Define the binary digits $(a^2)_i$, $(ab)_i$, $(b^2)_i$, $(a^2+ab)_i$
and $(ab+b^2)_i$, by $\{a^2\}/n=\sum_{i=1}^\infty (a^2)_i \,2^{-i}$, etc.
Then the identical calculation which led to $(4.7a)$ gives us
$$\{\ell'_ia\}+\{\ell_i'b\}=\{2^ia^2\}+\{2^iab\}+\left\{\matrix{
n&{\rm if}\sp (a^2)_{i+1}=(ab)_{i+1}=0\cr 0&{\rm if}\sp (a^2)_{i+1}+
(ab)_{i+1}=1\cr -n&{\rm if}\sp (a^2)_{i+1}=(ab)_{i+1}=1\cr}\right.
;\eqno(4.9)$$
with similar expressions for $\ell_i''$ and $\ell_j'''$ (for $\ell''_i$,
replace $a^2$ and $ab$ in (4.9) with $ab$ and $b^2$; for $\ell_j'''$
replace them with $a^2+ab$ and $ab+b^2$). Moreover, when $I>1$, we know
$a_i=0=b_i$ for $1\le i<I$, and $a_i+b_i=1$ for $I\le i\le M+1$; also
$(4.7b$) tells us $(a+b)_i=0$ for all $1\le i\le M+1$, except $(a+b)_{I-1}
=1$. This means: $\ell_i',\ell_i''>n/2$ for $1\le i\le I-2$;
for each  $I-1\le i\le M$ either $\ell'_i<n/2<\ell''_i$ or $
\ell''_i<n/2<\ell'_i$; for $0\le j\le M$, $\ell'''_j>n/2$, except for
$\ell'''_{I-2}<n/2$. From $(4.1b)$, these inequalities tell us how
the quantities like (4.9) compare to $n$, for all $1\le i\le M$ and
$0\le j\le M$.

For $I=1$, the identical inequalities hold for $\ell'_i,\ell''_i$,
except with $I$ there replaced with $I'$ here. The same applies to
$\ell'''_j$, except for the additional change that $\ell_0'''<n/2$.

Consider now $I\ge 3$. Since $a<n/2$, we have $\{a^2+ab\}=\{a^2\}+\{ab\}$;
since $b<n/2$ we have $\{ab+b^2\}=\{ab\}+\{b^2\}$. Note that induction on
$i$ gives us
$$(a^2+ab)_i+(ab+b^2)_i=1\quad {\rm for\ all}\ 1\le i\le I-2 \eqno(4.10a)$$
(both equal to 0 would contradict $\{a^2+ab\}+\{ab+b^2\}=n+3+\{ab\}>n$;
both 1 would contradict
$(4.1b$) with $\ell_{i-1}'''$). Then $(4.1b)$ with $\ell'''_{I-2}$ forces
$$(a^2+ab)_{I-1}=(ab+b^2)_{I-1}=1, \eqno(4.10b)$$
because the alternative, namely $(a^2+ab)_{I-1}+(ab+b^2)_{I-1}=1$, leads to
$$n>\{2^{I-2}(a^2+ab)\}+\{2^{I-2}(ab+b^2)\}=2\{2^{I-3}(a^2+ab)\}-n+2\{2^{I-3}
(ab+b^2)\}>2n-n=n,\eqno(4.10c)$$
a contradiction (the first inequality is the $\ell'''_{I-2}$ condition,
the second is $\ell'''_{I-3}$). And now, by the identical argument which
gave us $(4.7a$), we find that there exists an $I_0\ge I$ such that
$$n+3+\{ab\}=\{a^2+ab\}+\{ab+b^2\}=n+{n\over 2^{I_0-1}}+\eps,\quad 0<\eps
<{n\over 2^{M+1}}<2.\eqno(4.10d)$$
($I_0$ is the first index $\ge I$ such that $(a^2+ab)_{I_0}+(ab+b^2)_{I_0}
\ge 1$ -- $I_0\le M+1$ since otherwise $n+3<\{a^2+ab\}+\{ab+b^2\}<n+
{n\over 2^{M+1}}+{n\over 2^{M+1}}<n+4$, an impossibility.)

We can now fix $I$. To do this, note that the $\ell'_i$ conditions
tell us that either
$$\eqalignno{\{a^2\}+\{ab\}=&{n\over 2^{I_1-1}}+\eps' \quad {\rm for}\quad
 0<\eps' <{n\over 2^{I-1}},&(4.11a)\cr
{\rm or}\ \{a^2\}+\{ab\}=&{n\over 2}+{n\over 2^{I_1-1}}+\eps'\quad {\rm for}
\quad 0<\eps'<{n\over 2^{I-1}},&(4.11b)\cr}$$
for some $I_1>1$ in $(4.11a)$ and $I_1>2$ in $(4.11b)$ (if $(a^2)_i+(ab)_i
\ge 1$ for some $1<i\le I-1$, the derivation is identical to that of (4.7),
using $\ell'_i$ in place of $\ell_i$; otherwise $\{a^2\},\{ab\}<n/2^{I-1}$,
so (4.11) will be satisfied for some $I_1\ge I$). The identical
expressions apply to $\{ab\}+\{b^2\}$, of course, using $\ell''_i$ -- call
its parameters $I_2$ and $\eps''$. But according to (4.11), $(4.10b$) can
be satisfied iff $I_1=I=I_2$. But then (4.11) says $(4.10a$) can be satisfied
iff there are no $i$ between 1 and $I-1$ -- \ie $I=3$.

So $I=3$ is forced. $(4.7a)$ then reads $a+b=n/4+\eps'''$, where $0<\eps'''<2$
is fixed by $a+b$ being even. From $3+\{ab\}\equiv (a+b)^2$ (mod $n$) we get
$$3+\{ab\}=\left\{\matrix{(7n+18)/8&{\rm if}\ n\equiv 2\ ({\rm mod}\ 16)\cr
(5n+2)/8&{\rm if}\ n\equiv 6\ ({\rm mod}\ 16)\cr
(3n+18)/8&{\rm if}\ n\equiv 10\ ({\rm mod}\ 16)\cr
(n+2)/8&{\rm if}\ n\equiv 14\ ({\rm mod}\ 16)\cr}\right. .\eqno(4.12)$$
$(4.10d)$ is compatible with (4.12) only if $n\equiv 14$. In this case we
can compute $\{b^2\}$ as we did in Claim 6, and we find $\{b^2\}={1\over 8}n
+{b\over 2}+{11\over 4}$ (if $b\equiv 1$ (mod 4)) or ${5\over 8}n+{b\over 2}
+{11\over 4}$ (if $b\equiv 3$ (mod 4)).  In either case (at least for $n>14$),
 taking $\ell''_2$ gives us $n-11+2b+11>n$ , which contradicts $(4.1b)$
($I=3$ here, so $\ell''_2<n/2$).

Finally, consider the remaining possibility: $I=1$ and $M>I'\ge 3$.
$\ell'''_0<n/2$, but $(ab+b^2)_1=0$ by (4.8), so $(1+ab)_1=1$, \ie
$\{ab\}+1>n/2$, so $\{ab\}\ge n/2+2$ (it must be odd, and coprime to $n$).

Define $J>1$ by $n/2+n/2^J<\{ab\}<n/2+n/2^{J-1}$, and suppose for
contradiction that $J<I'$. Then as in the $(4.7b$) derivation (with $\ell'_i$
in place of $\ell_i$) we get $1+\{ab\}=n/2 +n/2^{J-1}+\eps$, $0<\eps<1$.
That is, $(a^2+ab)_i=0$ for all $i\le M+1$, except for $i=1$ and
$i=J-1$. But by (4.8), $(ab+b^2)_i=0$ for all $i\le M$. Now,
$\ell'''_{I'-2}<n/2$ produces a contradiction in $(4.1b)$:
$(a^2+ab)_{I'-1}=0=(ab+b^2)_{I'-1},$ since $J<I'$ by hypothesis.

Thus $I'\le J$. If we had $\{b^2\}>n/2$, then this would give us $n+2=
\{ab\}+\{b^2\}>n/2+2+n/2= n+2$, a contradiction.

Therefore $(b^2)_1=0$. As in $(4.10a)$, the constraints $\{ab\}+\{b^2\}>n$
and $(4.1b)$
with $\ell''_{i-1}$ tell us $(ab)_i+(b^2)_i=1$ for $i=2,\ldots,I'-1$. As in
$(4.10b)$, we also get
$(ab)_{I'}=(b^2)_{I'}=1$. Now $n+2=\{ab\}+\{b^2\}$ implies $(ab)_i=
(b^2)_i=0$ for $I'<i\le M$. Also, $(a^2)_i=0$ for all $i\le M+1$. But
either $\ell'_{I'}$ or $\ell''_{I'}$ will be less than $n/2$ -- whichever is
will violate $(4.1b)$.\qquad {\bf QED} to Prop.1 for $n\equiv 2$.

\bigskip\bigskip\noindent{{\bf Section 5. The simple-current chiral extension}}
\bigskip

In this section we use Thm.2 to find the possible values
$M_{\rho\mu},M_{\la\rho}$ for most $k$. We will find that except for four
values of $k$ considered in the next section,
a physical invariant will either be a permutation invariant, or an
automorphism of the {\it simple-current chiral extension}. The former are
listed in Thm.1; the latter are given in Thm.3 below. This will complete the
classification of $SU(3)_k$ for all $k\ne 5,9,21,57$.

\medskip\noindent{\bf Claim 7.}\quad Let $M$ be a level $k=n-3$ physical
invariant. Then for each $\la\in\p^k$, $M_{\rho\la}=M_{\la\rho}\in\{0,1\}$.
${\cal R}(M)\equi\R^\rho_L(M) =\R^\rho_R(M)$ will equal one of the following
sets:

\item{(a)} for $n\equiv 1,2$ (mod 3), $n\ne 8$, ${\cal R}(M)$ will equal $
\{\rho\}$;

\item{(b)} for $n\equiv 0$ (mod 3), $n\ne 12,24$, ${\cal R}(M)$ will equal
either $\{\rho\}$ or $\O_0\rho=\{\rho,A\rho,A^2\rho\}$;

\item{(c)} for $n=8$, ${\cal R}(M)$ will either equal $\{\rho\}$ or
$\{\rho,\rho'\}$, where $\rho'=(n/2-1,n/2-1)$;

\item{} for $n=12$, ${\cal R}(M)$ will either equal $\{\rho\}$, $\O_0\rho$,
or $\O_0\rho\cup\O_0\rho'$;

\item{} for $n=24$, ${\cal R}(M)$ will either equal $\{\rho\}$, $\O_0\rho$,
or $\O_0\rho\cup\O_0\rho'\cup\O_0\rho''\cup\O_0\rho'''$, where $\rho''=
(5,5)$ and $\rho'''=(7,7)$.

\noindent{\it Proof for $n\ne 8,12,24,60$} \quad We will defer the proof of
Claim 7 for the heights $n=8,12,24,60$ to the next section.

Claim 7 is automatic for $n\equiv 1,2,5,7,10,11$ (mod 12), by Thm.2 and
(P3). For the other levels, we will show $\R^\rho_L(M)$ must
equal one of the given possibilities, and also that each $M_{\la\rho}\in
\{0,1\}$. By symmetry the same comments apply to $\R^\rho_R(M)$ and
$M_{\rho\la}$, so $M_{\rho\la}=M_{\la\rho}$ and $\R^\rho_L(M)=\R^\rho_R(M)$
follow from Lemma 1(d).

Consider $n\equiv 4,8$ (mod 12). Thm.2 tells us $\R^\rho_L\subseteq
\{\rho,\rho'\}$. $\rho=\rho'$ for $n=4$, so we may assume $n> 8$ here.
We may assume $1\le m\equi M_{\rho'\rho}$ (otherwise $\R^\rho_L(M)=\{\rho\}$
and we are done). Then (see Lemma 1(a))
$$\eqalignno{
0\le& \,s^{(1,2)}_L(M)=S^{(n)}_{(1,2),\rho}\cdot 1 +S^{(n)}_{(1,2),\rho'}\cdot
m&\cr
=&\,{2\over \sqrt{3}n}\bigl\{(1+m)\sin[2\pi/ n]+(1-m)\sin[4\pi/ n]
-(1+m)\sin[6\pi/ n]\bigr\}&\cr
\le&\, {2\over \sqrt{3}n}(1+m)\bigl\{\sin[2\pi/n]-\sin[6\pi/n]\bigr\}.
&(5.1)\cr}$$
For $n>8$, the RHS of $(5.1$) is negative. Therefore, for $n>8$ $M_{\rho'\rho}
=0$ so $\R^\rho_L(M)=\{\rho\}$.

Now suppose $n\equiv 0$ (mod 3), and $\{\rho\}\ne\R^\rho_L(M)\subseteq \O_0
\rho$. Then Lemma 1(b) says $M_{A\rho,\rho}\ne 0$ iff $M_{A^2\rho,\rho}\ne 0$,
so both must be non-zero. Lemma 1(c) now tells us $1=M_{\rho\rho}=
M_{A\rho,\rho}=M_{A^2\rho,\rho}$, so $\R^\rho_L(M)=\O_0\rho$.

All that remains is $n\equiv 0$ (mod 12), where Thm.2 tells us $\R^\rho_L(M)
\subseteq\O_0\rho\cup\O_0\rho'$.
Define $m=\sum_{i=0}^2M_{A^i\rho,\rho}$, and $m'=\sum_{i=0}^2M_{A^i\rho',\rho}$
-- we may suppose
$m'\ge 1$ (otherwise $\R^\rho_L(M)\subseteq\O_0\rho$, which was done in the
previous paragraph). We will first prove that $m\le m'$. There are two cases
to consider: by Lemma 1(b) either $\J_L=\{A^0\}$ or $\{A^0,A,A^2\}$. In the
first case $m=1$ by (P3), so by supposition $m\le m'$. For the second case,
Lemma 1(c) tells us $m=3M_{\rho\rho}=3$ and $m'=3M_{\rho'\rho}\ge 3$, so again
$m\le m'$.

Then, by Lemma 1(a) again and by $(1.6d$), we have
$$\eqalignno{
0\le &\,s^{(1,4)}_L(M)=S^{(n)}_{(1,4),\rho}\cdot m+S^{(n)}_{(1,4),\rho'}\cdot
m' &\cr =&\,{2\over \sqrt{3}n}\bigl\{(m+m')\sin[2\pi/n]+(m-m')\sin[8\pi/n]
-(m+m')\sin[10\pi/n]\bigr\}&\cr
\le &\,{2\over \sqrt{3}n}(m+m')\bigl\{\sin[2\pi/n]-\sin[10\pi/n]\bigr\}.
&(5.2)\cr}$$
But the RHS of (5.2) will be negative unless $n=12$.\qquad QED\medskip

When ${\cal R}(M)=\{\rho\}$, Lemma 2 tells us $M$ will be listed in Thm.1. We
will handle the anomolous $\rho$-couplings of $n=8,12,24$, or 60 in Sect.6;
here we consider $n\equiv 0$ (mod 3) and ${\cal R}(M)=\O_0\rho$ -- this
corresponds
to a simple-current chiral extension. But first let us review what we know.

Recall the definition of {\it triality}: $t(\la)\equiv
\la_1-\la_2$ (mod 3). Let $\p_0$ denote the set of all weights $\la\in\p^k$
with $t(\la)\equiv 0$, and let $\p_{[]}=\p_0/\O_0$ be the set of all
orbits $[\la]=\O_0\la\subset\p_0$. Lemma 1(e) tells us $\p_L(M)=
\p_R(M)=\p_0$. Note that there is only one `fixed point'
of $\O_0$, namely $f=(n/3,n/3)$. Lemma 3(c) defines a mapping $\sigma$
with a domain and range contained in $\p_{[]}\sim \{f\}$, with the
property that when $\la,\mu\ne f$, then $M_{\la\mu}\ne 0$ iff $M_{\la\mu}=1$
iff $[\mu]=\sigma[\la]$. So we
already know a considerable amount about $M$. All that remains is to
understand what $M$ looks like at the fixed point $f$, and then to
find $\sigma$. Write $\sigma'$ for the inverse of $\sigma$.
That $f$ can cause complications is apparent by looking
at the exceptional ${\cal E}_9^{(2)}$.

\medskip\noindent{\bf Theorem 3.} Suppose ${\cal R}(M)=\O_0\rho$. Then $M$
is either
the {\it simple-current invariant} ${\cal D}_k$ given in eq.(1.7$c$), its
conjugation ${\cal D}_k^c$, the exceptional ${\cal E}_9^{(2)}$ given
in $(1.7f)$, or its conjugation ${\cal E}_9^{(2)c}$.

\noindent{\it Proof}  \quad{\bf Step 1} We begin by investigating $M_{f\la},
M_{\la f}$.

Look at the decompositions $\oplus_i B_i$ of
$MM^T$, and $\oplus_j B_j'$ of $M^TM$ (see (2.4$a$)), where $B_1,B_1'$
`contain' $\rho$, and $B_\ell,B_\ell'$ `contain' $f$. Now, $B_1=B_1'=
B_{(3,3)}$ (see (2.4$d$)), so by Lemma 3(b) $r(B_\ell)=r(B'_\ell)=9$.
One thing this means is (see ($2.4c$))
$$\eqalignno{\bigl(B_\ell\bigr)_{ff}=&\sum_{\mu}M_{f\mu}^2=M_{ff}^2+3
\sum_{[\mu]\ne [f]}M_{f\mu}^2\le 9,&(5.3a)\cr
\bigl(B_\ell'\bigr)_{ff}=&\sum_{\la}M_{\la f}^2=M_{ff}^2+3
\sum_{[\la]\ne [f]}M_{\la f}^2\le 9,&(5.3b)\cr}$$
with equality in $(5.3a$) iff $B_\ell$ is $1\times 1$, and with equality
in $(5.3b$) iff $B_\ell'$ is $1\times 1$. Let $[\la^1],[\la^2],\ldots$ denote
the different orbits $[\la]\subset\R^f_L(M)\sim\{f\}$, and $[\mu^1],[\mu^2],
\ldots\subset\R^f_R(M)\sim\{f\}$. From (5.3) we read off that $M_{\la^i f}
=1=M_{f\mu^j}$ for all $i,j$. We also see that there are precisely six
possibilities for $\R^f_L(M)$ (similarly for $\R^f_R(M)$): $\R^f_L(M)$
will equal either $[f]$, $[\la^1]$, $[f]\cup[\la^1]$, $[\la^1]\cup[\la^2]$,
$[f]\cup[\la^1]\cup[\la^2]$, and $[\la^1]\cup[\la^2]\cup[\la^3]$. Each
$\la^i\ne f$, and lie on distinct orbits. Call these cases 10, 01, 11, 02,
12, and 03, respectively.

To begin note that $\R^f_L(M)$ and $\R^f_R(M)$ are always
of equal {\it case} (though {\it a priori} they could be unequal as sets).
To see this, let their cases be $ab$ and $cd$, respectively, where $a,c=0,1$
and $b,d=0,1,2,3$. Now, $a=c$ because $a=0$ iff $M_{ff}=0$ iff $c=0$. Also,
$b=d$ because the cardinality of the domain of $\sigma$, namely $(n-1)(n-2)/2
-3b-1$, must equal the cardinality of the domain of $\sigma'$, namely
$(n-1)(n-2)/2-3d-1$.

Case 10: Note that $B_\ell$ is $1\times 1$. Therefore equality must
hold in $(5.3a$), which for case 10 requires $M_{f\la}=M_{\la f}=3
\delta_{\la f}$.

Case 0b, $b=1,2,3$:  Note that for any case $0b$, $B_\ell$ will be $1\times 1$.
Therefore we must have $9=3b$, since $M_{f\mu^i}=1$. This means that of
these, only case 03 works.

Case 11: The question here is, what can $M_{ff}$ equal? Eq.(5.3$a$)
demands it be 1 or 2. Note that
$$B_\ell=\left(\matrix{M^2_{ff}+3&M_{ff}\cr M_{ff}&1 \cr}\right),
\eqno(5.3c)$$
again using the fact that $M_{f\mu^1}=M_{\la^1f}=1$. This must have
$r(B_\ell)=9$; we find that this requires $M_{ff}=2$.

Case 12: The same reasoning as used in case 11 applies here. We find
that $M_{ff}=1$ is forced here.

{\bf Step 2} Now, look at the commutation relations that $\sigma$ and
$\la^i$ must satisfy. First, choosing $\mu\not\in\R^f_R(M)$, and any
$\la^i\ne f$ lying in $\R^f_L(M)$, we get
$$S^{(n)}_{f\mu}=\sum_{\nu\in\p^k} M_{\la^i\nu}\,S^{(n)}_{\nu\mu}=
\sum_{\nu\in\p^k}
S^{(n)}_{\la^i\nu}\,M_{\nu\mu}=3S^{(n)}_{\la^i,\sigma'\mu},\eqno(5.4a)$$
where by ``$\sigma'\mu$'' in $(5.4a$) we mean any element of $\sigma'[\mu]$
-- because $\la^i\in\p_0$ $(1.6d)$ tells us the RHS of $(5.4a$) is unaffected
by this choice.

We may assume $n>6$ (for $n=6$, $(2.1c$) alone is enough to
force the partition function to be ${\cal D}_3={\cal D}_3^c$).
Thus $(2,2)\ne f$. Now
$(2,2)\in\R^f_L(M)$ only if $3S^{(n)}_{(2,2),\rho}=S^{(n)}_{f \rho }$ using
$(5.4a)$, \ie only if
$$\sin({2\pi\over n})\,\sin({2\pi\over n})\,\sin({4\pi\over n})={\sqrt{3}
\over 8}, \eqno(5.4b)$$
using $(2.1a$). But the LHS is a strictly decreasing function of $n\ge 9$,
so equality there can only happen for one value of $n$. It turns out $n=12$
is this value. Therefore, for $n\ne 12$, $(2,2)\not\in\R^f_L(M)$ and $(2,2)
\not\in\R^f_R(M)$.

We would like to show that in fact $\sigma[(2,2)]=[(2,2)]$. If $n$ is odd,
we get that
easily: $2\in\C_{3n}$ so Lemma 4(a) tells us $1=M_{11,11}=M_{22,22}$.
For $n$ even, write $\sigma([2,2])=[(a,b)]$ -- we may choose $a\le b\le
n-a-a$. $(MS)_{22,\rho}=(SM)_{22,\rho}$ gives us
$$\sin({a\pi\over n})\,\sin({b\pi\over n})\,\sin({(a+b)\pi\over
n})=\sin({2\pi\over n})\,\sin({2\pi\over n})\,\sin({4\pi\over n}).\eqno(5.4c)$$
Then the LHS of $(5.4c$) will be larger than the RHS, unless either
$a=b=2$, or $a=1$, or $n-a-b<4$. The latter possibility requires
$a=b=3$ and $n=9$, \ie $(2,2)\in\R^f_L(M)$, which was ruled out already.
$a=1$ fails $(2.1c$), for $n$ even: $1+b+b^2$ is odd, but
$4+4+4$ is even, so they cannot be equal (mod $3n$). Thus for all $n$
(except possibly $n=12$), we have $\sigma[(2,2)]=[(2,2)]$.

{\bf Step 3} Suppose for contradiction that $\la^i\in\R^f_L(M)$ for some
$\la^i\ne f$. Write $\la^i=(a,b)$. Then $(5.4a)$ with $\mu=\rho$ and
$\mu=(2,2)$, respectively, tells us
$$\eqalignno{\sin({a\pi\over n})\,\sin({b\pi\over n})\,\sin({(a+b)\pi\over
n})=&{\sqrt{3}\over 8},&(5.5a)\cr
\sin({2a\pi\over n})\,\sin({2b\pi\over n})\,\sin({2(a+b)\pi\over
n})=&{-\sqrt{3}\over 8}.&(5.5b)\cr}$$
Dividing these equations gives
$$\cos({a\pi\over n})\,\cos({b\pi\over n})\,\cos({(a+b)\pi\over
n})={-1\over 8}.\eqno(5.5c)$$
We claim that in the interval $0<a,b,a+b<n$, the LHS of $(5.5c$) has a
global minimum of $-1/8$, and it achieves that only at $a=b=n/3$.
This can be trivially proved by differentiating: the derivative with
respect to $a$ is proportional to $\cos(b\pi/n)\,\{\sin(a\pi/n)\,\cos((a+b)\pi
/n)+\cos(a\pi/n)\,\sin((a+b)\pi/n)\}=\cos(b\pi/n)\,\sin((2a+b)\pi/n)$,
etc.

In other words, there is a {\it unique} solution to $(5.5c$): $a=b=n/3$. But
this is not a solution to $(5.5a$). Therefore $(5.5a,b$) are incompatible,
so no such $\la^i$ can be found.

What we have shown is that whenever $(2,2)\not\in\R^f_L(M)$ (\eg $n\ne 12$),
then $\R^f_L(M)=\R^f_R(M)=[f]$, so $M_{f\la}=M_{\la f}=3\delta_{\la f}$.
Defining $\sigma[f]=[f]$ thus makes $\sigma$ into a permutation
of $\p_{[]}$.

{\bf Step 4} Choose any $\la,\mu\in\p_0$. Expanding out $(S^{(n)}M)_{\la\mu}
=(M\,S^{(n)})_{\la\mu}$ gives us $3S^{(n)}_{\la,\sigma'\mu}=3
S^{(n)}_{\sigma\la,\mu}$, \ie
$$S^{(n)}_{\la\mu}=S^{(n)}_{\sigma\la,\sigma\mu}. \eqno(5.6)$$

So $\sigma$ is very much like a permutation invariant, and we can hope the
argument in Sect.3 carries over with minimal change. To some extent this
is true, but we must be careful here because $\sigma$ is only defined on
$\p_{[]}$ -- the $\sigma$ of a true permutation invariant would act on $\p^k$.

In Sect.3 we look at $N^{(n)}_{\la\la\la}$. This doesn't work as well here
(see eq.(5.10) in \SU). Instead, we will use $N^{(n)}_{\la ff}$. From (3.3)
we compute:
$$N^{(n)}_{\la ff}=\left\{\matrix{{\rm min}\{\la_1,\la_2,n-\la_1-\la_2\}&
{\rm if}\sp \la\in\p_0\cr 0&{\rm otherwise}\cr}\right. .\eqno(5.7a)$$
Now, from Verlinde's formula $(3.2a$), and eq.(5.6), and the fact that
$S^{(n)}_{\mu f}=0$ if $\mu\not\in\p_0$ (this follows from $(1.6d)$ and
$S^{(n)}_{\mu f}=S^{(n)}_{\mu, Af}$), we see that $N^{(n)}_{\la ff}=
N^{(n)}_{\sigma\la, ff}$ $\forall\la\in\p_0$. Thus
$$\{\la_1,\la_2,n-\la_1-\la_2\}\cap\{(\sigma\la)_1,(\sigma\la)_2,n-(\sigma
\la)_1-(\sigma\la)_2\}\ne \{\}.\eqno(5.7b)$$
(5.6) also gives us $S^{(n)}_{\rho\la}=S^{(n)}_{\rho,\sigma\la}$ for all
$\la\in\p_0$. Therefore Claim 1 applies, and we get
$$\sigma[\la]=[\la]\quad{\rm or}\quad [C\la],\quad \forall \la\in\p_0.
\eqno(5.7c)$$
\noindent{\bf Claim 8.} For $\la\in\p^k$, $S^{(n)}_{(1,4),\la}
=S^{(n)}_{(1,4),C\la}$ iff $[\la]=[C\la]$.

\noindent{\it Proof} \quad The proof is very similar to that of Claim 2.
Suppose first that $S^{(n)}_{14,1a}=S^{(n)*}_{14,1a}$ (as in Claim 2 we
need not limit ourselves to integral $n=k+3$ or $a$). Then we get
$${\tilde c}_n(3a+2)+{\tilde c}_n(2a-1)+{\tilde c}_n(a+3)={\tilde c}_n(2a+3)
+{\tilde c}_n(a-2)+{\tilde c}_n(3a+1),\eqno(5.8a)$$
where ${\tilde c}_n(x)=\cos(2\pi x/n)$. Putting $b=a+{1\over 2}$ we get
$${\tilde p}(b,n)={\tilde c}_n(3b+{1\over 2})+{\tilde c}_n(2b-2)+{\tilde c}_n
(b+{5\over 2})-{\tilde c}_n(3b-{1\over 2})-{\tilde c}_n(2b+2)-{\tilde c}_n
(b-{5\over 2}).\eqno(5.8b)$$
Then as before ${\tilde p}(b,n)={\tilde s}_n(b)\,p^{(n)}_2({\tilde c}_n(b))$,
where $p^{(n)}_2$ is a quadratic polynomial. Thus ${\tilde p}(b,n)$ will
have at most 2+2 zeroes (2 from ${\tilde s}_n$, and $\le 2$ from $p^{(n)}_2$)
in the range $0\le b\le n/2$. But $b=0,{1\over 2},{3\over 2},{n\over 2}$ are
clearly zeroes. Therefore as long as $n\ge 3$, there are no solutions $a$
to $S^{(n)}_{14,1a}=S^{(n)*}_{14,1a}$ in the range $1<a<{n\over 2}-{1\over 2}$.

Now choose any $\la\in\p^k$. Then $\exists \la'\in\O\la$ such that $\la'_1\le
\la_2'\le n-\la_1'-\la_2'$. Put $a=\la_2'/\la_1'$, $n'=n/\la_1'$, so
$1\le a\le {n'\over 2}-{1\over 2}$ and $n'\ge 3$. We find that
$S^{(n)}_{14,\la}=S^{(n)*}_{14,\la}$ iff $S^{(n)}_{14,\la'}=S^{(n)*}_{14,
\la'}$  iff $S^{(n')}_{14,1a}=S^{(n')*}_{14,1a}$
iff $a=1$ or $a={n'\over 2}-{1\over 2}$. $a=1$ corresponds to $\la'_1=\la_2'$,
\ie $C\la'=\la'$. $a={n'\over 2}-{1\over 2}$ corresponds to $\la_2'=n-\la_1'
-\la_2'$, \ie $CA^2\la'=A^2\la'$. \qquad QED\medskip

The conclusion to the proof of Thm.3 is as
in Sect.3. From $(5.7c$) we may assume, by conjugating if necessary, that
$\sigma[(1,4)]
=[(1,4)]$. Suppose $\sigma[\la]=[C\la]$ for some $[\la]\in\p_{[]}$. Then
(5.6) tells us
$S^{(n)}_{(1,4),\la}=S^{(n)}_{(1,4),C\la}$, so by Claim 8 $[\la]=[C\la]$.
Therefore $\sigma[\la]=[\la]$ for all $\la\in\p_0$, so $M$ is the invariant
${\cal D}_k$.\qquad QED to Thm.3

\bigskip\bigskip\noindent{{\bf Section 6. The exceptional levels}}\bigskip

The analysis in the last section avoided four heights: $n=8$, 12,
24 and 60. There are various ways we can handle these. One way is the
lattice method, employed in \GH. But this is a bit of overkill: that
method finds much more than just the physical invariants. The consequence
is that, for higher levels (or higher ranks), the lattice method becomes
unfeasible. E.g. it has never been worked out for $n=60$.

The methods developed in this paper however work for these exceptional
levels -- they merely require a bit more effort. That will be the task
in this section: to complete the $SU(3)_k$ classification for these four
levels. The first thing to do is to complete the proof of Claim 7 for
these levels.

\medskip
\noindent{\it Proof of Claim 7 for $n=8,12,24,60$}\quad First consider $n=8$.
Again put $m=M_{\rho'\rho}$. Then as in (5.1) we get
$$0\le s^{(1,2)}_L(M)={2\over \sqrt{3}n}(1-m).\eqno(6.1a)$$
Therefore either $m=0$ or $m=1$ (\ie $\R^\rho_L(M)=\{\rho\}$ or $\{\rho,
\rho'\}$, respectively).

Next consider $n=12$. Again put $m=\sum_{i=0}^2M_{A^i\rho,\rho}$ and $m'=
\sum_{i=0}^2M_{A^i\rho',\rho}$. $m'=0$ was done in Sect.5, so consider here
$m'>0$. Then as before, $m'\ge m$ and $m=1$ or 3. Eq.(5.2) becomes
$$0\le s^{(1,4)}_L(M)={2\over \sqrt{3}n}(m-m')({\sqrt{3}\over 2}).\eqno(6.1b)$$
Therefore $m'=m=1$ or 3. If $m=3$ then $M_{\la\rho}=0$ or 1 for all $\la$,
by Lemma 1(c), and $\R^\rho_L(M)=\O_0\rho\cup\O_0\rho'$. But if $m=1$, only
one $\la\in\O_0\rho'$, say $\la=A^\ell \rho'$, has $M_{\la\rho}\ne 0$. Then
we have $M_{A^i\rho',\rho}=\delta_{i\ell}$. To show this is impossible look
now at $s_L^{(1,2)}(M)$: depending on the value of $\ell$ this will either
be non-real ($\ell=1,2$) or negative $(\ell=0$).

Now consider the more difficult case $n=24$. Write $\rho''=(5,5)$, $\rho'''=
(7,7)$, $m=\sum_{i=0}^2M_{A^i\rho,\rho}$, $\ldots, m'''=\sum_{i=0}^2 M_{A^i
\rho''',\rho}$. We may restrict ourselves to the case where $m''>0$ or
$m'''>0$,
since $m''=m'''=0$ was done in Sect.5. As usual, $m=1$ or 3, and either
$m'=0$ or $m'\ge m$, either $m''=0$ or $m''\ge m$, and either $m'''=0$ or
$m'''\ge m$. Write $s(x)$ here for $\sin(\pi x/12)$. Then
$s^{(2,2)}_L(M)$, $s^{(3,3)}_L(M)$ and $s^{(4,4)}_L(M)$ give us
$$\eqalignno{0\le & \,2(m-m'+m''-m''')\,s(2)-(m-m'-m''+m''')\,s(4),&(6.2a)\cr
0\le & \,2(m+m'-m''-m''')\,s(3)-(m-m'+m''-m''')\,s(6),&(6.2b)\cr
0\le & \,2(m-m'-m''+m''')\,s(4)-(m-m'-m''+m''')\,s(8),&(6.2c)\cr}$$
respectively. These give us
$$\eqalignno{0\le & \,m-m'-m''+m''',&(6.3a)\cr
0\le & \,m-m'+m''-m''',&(6.3b)\cr
0\le & \,m+m'-m''-m'''.&(6.3c)\cr}$$
In particular, $(6.3a)$ comes immediately from $(6.2c)$; $(6.3b)$ comes
from $(6.2a)$ and $(6.3a)$; and $(6.3c$) comes from $(6.2b)$ and $(6.3b$).

Now adding eqs.$(6.3a,b)$, eqs.($6.3a,c)$, and eqs.$(6.3b,c)$ give us
$m'\le m$, $m''\le m$,  and $m'''\le m$. So each $m',m'',m'''$ will either
equal 0 or $m$. If $m''=m$, then ($6.2b$) tells us $m'=m'''=m$; if
$m'''=m$, then (6.2$a$) tells us $m'=m''=m$. Thus $m=m'=m''=m'''=1$ or 3.

As in the $n=12$ case, $m=3$ leads to the exceptional $\rho$-coupling given in
Claim 7. If $m=1$, there are numbers $\ell'$, $\ell''$, $\ell'''$ such that
$M_{A^i\rho',\rho}=\delta_{i\ell'}$, $M_{A^i\rho'',\rho}=\delta_{i\ell''}$, and
$M_{A^i\rho''',\rho}=\delta_{i\ell'''}$. This leads to 27 possibilities,
all of which fail the $s^{(3,2)}_L(M)\ge 0$ test, as can be easily
checked on a computer.

Finally, consider $n=60$. The reasoning and calculations are very similar
to that for $n=24$. Here, put $\rho''=(11,11)$, $\rho'''=(19,19)$, and
define $m,m',m'',m'''$ as before. Looking at $s^{(3,3)}_L(M)$, $s^{(6,6)}_L(M)$
and $s^{(10,10)}_L(M)$ give us eqs.(6.3) again. As before, we can force
$m=m'=m''=m'''$ (use $s^{(3,3)}_L(M)$ and $s^{(5,5)}_L(M)$). But the difference
here is that $s^{(2,5)}_L(M)<0$, so neither $m=1$ nor $m=3$ work. \qquad
QED to Claim 7\medskip

Our next task is to handle the remaining case of Thm.3: $n=12$, so $f=(4,4)$,
 and $M_{22,44}=
1$. First, on a computer we can find for each $\la=(a,b)\in\p_0$ all $\mu
=(a',b')\in\p_0$ satisfying $(2.1c$) as well
as the parity rule Lemma 4(b). We find any $\la,\mu\in[1,1]\cup[5,5]$ satisfy
these two conditions, as do any $\la,\mu\in[2,2]\cup[4,4]$, $\la,\mu\in[3,3]$,
and $\la,\mu\in[1,4]\cup[4,1]$. This means $M_{\la f}\ne 0$ only if
$\la\in[2,2]\cup[f]$, so we find that $M$ is ``case 11'' here (see Step 1
of the proof of Thm.3). Therefore $M_{ff}=2$, and we know all values of
$M_{\la f},M_{f\la}$.

All we need to do is determine $\sigma[5,5]$, $\sigma[3,3]$, $\sigma[1,4]$
and $\sigma[1,4]$. Eq.($2.1c$) and the parity rule tell us $\sigma[3,3]=[3,3]$.
That $\sigma[5,5]=[5,5]$ can be seen by Lemma 4(a) with $\ell=5$ applied to
$M_{11,11}$. Now
note that $M_{14,14}+M_{14,41}=1$ (\ie $\sigma[1,4]$ equals either
[1,4] or [4,1]). Conjugating if necessary, we may assume $M_{14,14}=1$. Then
Lemma 4(a) with $\ell=5$ tells us $M_{41,41}=1$. This determines all unknown
matrix elements $M_{\la\mu}$, and we find we have $M={\cal E}^{(2)}_9$.

Finally, we must address the exceptional couplings listed in Claim 7. The idea
is to use the various results we have to reduce the number of possible $M$
to a small number which can then be checked on a computer for modular
invariance. Together, Thms.1,3 and 4 complete the SU(3) classification, for
any level $k$.

\medskip\noindent{\bf Theorem 4.}\quad (i) For $n=8$, if ${\cal R}(M)=\{\rho,
\rho'\}$ then $M$ either equals ${\cal E}_5$ or its conjugation.

\item{(ii)} For $n=12$, if ${\cal R}(M)=\O_0(1,1)\cup\O_0(3,3)$
then $M={\cal E}_9^{(1)}$.

\item{(iii)} For $n=24$, if ${\cal R}(M)=\O_0(1,1)\cup\O_0(5,5)
\cup\O_0(7,7)\cup\O_0(11,11)$ then $M={\cal E}_{21}$.

\noindent{\it Proof}\quad (i) We know from Claim 7 that $M_{\rho\la}=
M_{\la\rho}$, for all $\la$, and that these all vanish except for $\la=(1,1)$
and $\la=(3,3)$, for which $M_{\rho\la}=M_{\la\rho}=1$. Note that $(3\rho)^+
=(3,3)$, and $(3(3,3))^+=\rho$. Therefore Lemma 4(a) with $\ell=3$ applied
to $M_{11,11}=1$ gives us $M_{33,33}=1$. Also, $M_{33,\mu}=M_{11,(3\mu)^+}=0$
unless $\mu=\rho$ or (3,3) (same for $M_{\la,33}$). Thus, if we expand
$M$ as in $(2.4a$), with $B_1$ `containing' $\rho$ as usual, then $B_1=
B_{(1,2)}$, so Lemma 3(b) tells us $r(B_i)=2$ for all $i$, and each $M_{\la\mu}
\le 2$. We have reduced the number of possibilities for $M$ to a finite
number, and with a bit more effort can reduce this number further.

Lemma 1(a) tells us that $\p_L(M)=\p_R(M)=[1,1]
\cup[3,3]\cup\O[1,3]$. Again, $(2.1c$) and Lemma 4(b) strongly restrict
the possible couplings, and we find $M_{\la\mu}\ne 0$ only for
$\la,\mu\in [1,1]\cup[3,3]$, or for $\la,\mu\in\O[1,3]$. The relation
$S^{(n)}M=MS^{(n)}$ evaluated at $((6,1),\rho)$ tells us that $M_{61,16}+
M_{61,61}=M_{61,32}+M_{61,23}$=1; wlog (by conjugating if necessary) we may
suppose $M_{61,61}=1$. Then the familiar calculation (see $(2.3b)$)
$$1=M_{A\rho,A\rho}=\sum_{\la,\mu}S^{(n)*}_{\rho\la}\,M_{\la\mu}\,
S^{(n)}_{\mu\rho}\omega^{t(\mu)-t(\la)}\le M_{\rho\rho}=1 \eqno(6.4)$$
tells us $t(\la)\equiv t(\mu)$ (mod 3) whenever $M_{\la\mu}\ne 0$.

There is now a small number of possibilities for the remaining $M_{\la\mu}$.
Each can be easily checked on a computer for modular invariance (it is more
convenient to use here $SM=MS$ rather than $M=S^{\dag}MS$).

(ii) This is easier. Lemma 1(a) says $\p_L(M)=\p_R(M)=[1,1]\cup[3,3]\cup
[5,5]$. $(2.1c$) and Lemma 4(b) tell us $M_{\la\mu}\ne 0$ implies
$\la,\mu\in[1,1]\cup[5,5]$ or $\la,\mu\in[3,3]$. $\J_L(M)=\J_R(M)=\O_0$,
so Lemma 1(c) tells us the only independent parameters are $M_{11,11}=
M_{11,55}=M_{55,11}=1$, $M_{55,55}$ and $M_{33,33}$. $M_{55,55}=1$ by
Lemma 4(a) using $\ell=5$ and $M_{11,11}=1$, so (using usual notation)
$B_1=B_{(1,6)}$. But $B_2=B_{(M_{33,33},3)}$ so $3M_{33,33}=r(B_2)=r(B_1)=6$,
\ie $M_{33,33}=2$. We have derived $M={\cal E}^{(1)}_9$.

(iii) $\p_L(M)=\p_R(M)=[1,1]\cup[5,5]\cup[7,7]\cup[11,11]\cup[5,8]\cup[8,5]
\cup[1,7]\cup[7,1]$. $(2.1c$) tells us $M_{\la\mu}\ne 0$ implies either
$\la,\mu\in[1,1]\cup[5,5]\cup[7,7]\cup[11,11]$ or $\la,\mu\in[5,8]\cup
[8,5]\cup[1,7]\cup[7,1]$. Lemma 4(a) applied to $\ell=5,7,11$ and to
$M_{11,11}=M_{11,55}=M_{11,77}=M_{11,1111}=1$ tells us $M_{aa,bb}=1$ for
each choice $a,b=1,5,7,11$. Also, $\J_L(M)=\J_R(M)=\O_0$. Together with
Lemma 1(c), we find that $B_1=B_{1,12}$.

Note that from Lemmas 4(a) and 1(c), we can similarly deduce all the remaining
values of $M_{\la\mu}$ once we know the four values $M_{58,58},M_{58,85},
M_{58,17},M_{58,71}$. In fact we find from Lemmas 4(a) and 1(c) that
for each $\la\in[5,8]\cup[8,5]\cup[1,7]\cup[7,1]$, each row sum $\sum_\mu
M_{\la\mu}$ is equal (\ie independent of $\la$). Eq.(2.4$b)$ then tells
us these sums must equal $r(B_1)=12$, so $M_{58,58}+M_{58,85}+M_{58,17}+
M_{58,71}=4$. The small number of possibilities can be checked on a
computer for modular invariance. \qquad QED

\bigskip\bigskip\noindent{{\bf Conclusion}}\bigskip

In this paper we rewrote the classification in \SU{}, paying attention
to the four points mentioned in the abstract.

There have been additional modular invariant classifications since
\SU{} was written. For example, the {\it simple-current invariants} have been
classified \SCH. Among other things, \AA{} completes the classification for
$SU(2)_{k_1}\times SU(2)_{k_2}$, $\forall k_1,k_2$.
All heterotic invariants of small rank are now known \HET, as are many of the
$c=24$ meromorphic theories \MER. New classifications have recently been
found \GW{} for diagonal GKO cosets corresponding to SU(2) and SU(3).
Most physical invariants are `obvious' -- \ie either due to outer automorphisms
of the affine algebra, or to conformal embeddings. But new exceptionals are
always being found.

Future work along these lines include attempting to find all permutation
invariants for $SU(N)_k$ \RU, and the possible chiral algebras for \eg
$SU(2)_{k_1}\times\cdots\times SU(2)_{k_r}$ \ST. Both these will be very
important accomplishments. My own interests in this area lie in pursuing
classifications for $SU(N)_k$, say, for all `sufficiently large' $k$ -- there
are reasons for believing that many simplifications occur in the large
$k$ limit. In particular, I would like $k$ (or $n$) to play a more `dynamic'
role in the arguments.

But isolated classifications can be rather sterile. The interesting thing
is to see if they shed any new light on the subject, \eg disclose
new connections or possibilities involving other areas of math or physics.
Indeed this seems to be the case with these modular invariant classifications.
The most famous example is the A-D-E classification of the $SU(2)_k$
physical invariants \CIZ. This $SU(3)_k$ classification also hints of deep
interconnections with other areas. Indeed \ARSY{} has discovered (apparently
completely by accident! \RU) a fascinating connection involving Fermat curves
\KR. For me personally, this is the main motivation to continue in this area:
trying to understand a little better these mysteries.

There are many interconnections between modular invariants of different
algebras and levels. This can be seen in this paper. For example when $n\equiv
0$ (mod 4), the parity rule $\eps(\ell\la)=\eps(\ell\mu)$, $\ell\in\C_n$,
for $SU(3)_{n-3}$ is intimately connected with that for $SU(2)_{n/2-2}$ and
$SU(3)_{n/2-3}$ -- exactly how depends on the values of $\la_1,\la_2,\mu_1,
\mu_2$ (mod 2). We saw the $\mu=\rho$ case of this in the proof of Prop.1.
The `reason' for these connections is that $\ell\in\C_n$ iff $\ell+n/2\in\C_n$
for these $n$. It is natural to try to better understand and exploit these
various interconnections.

\bigskip \noindent{\it Acknowledgements.}
A number of people supported me at various stages of \SU, and this paper.
I would particularly like to thank Patrick
Roberts, who was my sounding board through this whole process and
could have had his name on \SU. I also benefitted throughout from
conversations with
Quang Ho-Kim and C.S. Lam, and more recently with Antoine Coste,
Claude Itzykson, Jean Lascoux, Philippe Ruelle, Yassen Stanev and Mark Walton.
The generosity of the IHES is also greatly appreciated.

\bigskip\bigskip \noindent{{\bf References}} \bigskip

\item{[1]} Altschuler, D., Lacki, J., Zaugg, Ph.: The affine Weyl group
and modular invariant partition functions. Phys.\ Lett.\ {\bf B205} 281-284
(1988)

\item{[2]} Bauer, M., Itzykson, C.: Modular transformations of SU(N)
affine characters and their commutant. Commun.\ Math.\ Phys.\ {\bf 127} 617-636
(1990)

\item{[3]} B\'egin, L., Mathieu, P., Walton, M.: $\hat{su}(3)_k$
fusion coefficients. Mod.\ Phys.\ Lett.\ A{\bf 7} 3255-3265 (1992)

\item{[4]} Bernard, D.: String characters from Kac-Moody automorphisms.
Nucl.\ Phys.\ {\bf B288} 628-648 (1987)

\item{[5]} Cappelli, A., Itzykson, C., Zuber, J.-B.: The A-D-E classification
of $A_1^{(1)}$ and minimal conformal field theories. Commun.\ Math.\ Phys.\
{\bf 113} 1-26 (1987)

\item{[6]} Christe, P., Ravanani, F.: $G_N\otimes G_L/G_{N+L}$ conformal
field theories and their modular invariant partition functions. Int.\ J.\ Mod.\
Phys.\ {\bf A4} 897-920 (1989)

\item{[7]} Coste, A., Gannon, T.: Remarks on Galois symmetry in RCFT.
Phys.\ Lett.\ {\bf B323} 316-321 (1994)

\item{[8]} Gannon, T.: WZW Commutants, Lattices, and Level-one
Partition Functions. Nucl.\ Phys.\ B{\bf 396} 708-736 (1993)

\item{[9]} Gannon, T.: The classification of affine SU(3) modular
invariant partition functions. Commun.\ Math.\ Phys.\ {\bf 161} 233-264 (1994)

\item{[10]} Gannon, T.: Towards a classification of
SU(2)$\oplus\cdots\oplus $SU(2) modular invariant partition functions.
IHES preprint P/94/21 (hep-th/9402074)

\item{[11]} Gannon, T., Ho-Kim, Q.: The low level modular invariant
partition functions of rank 2 algebras. Int.\ J.\ Mod.\ Phys.\ A (in press)
(hep-th/9304106)

\item{[12]} Gannon, T., Ho-Kim, Q.: The rank-four heterotic modular
invariant partition functions. IHES preprint P/94/5 (hep-th/9402027)

\item{[13]} Gannon, T., Walton, M.A.: On the classification of diagonal
coset modular invariants. (in preparation)

\item{[14]} Gantmacher, F.R.: The theory of matrices. Vol.II. New York:
Chesea Publishing Co. 1964

\item{[15]} Ka{\v c}, V.G.: Infinite Dimensional Lie Algebras, 3rd ed.
Cambridge: Cambridge University Press 1990

\item{[16]} Koblitz, N., Rohrlich, D.: Simple factors in the Jacobian
of a Fermat curve. Can.\ J.\ Math.\ {\bf XXX} 1183-1205 (1978)

\item{[17]} Kreuzer, M., Schellekens, A.N.: Simple currents versus
orbifolds with discrete torsion -- a complete classification. Nucl.\ Phys.\
B411 (1994) 97-121

\item{[18]} Moore, G., Seiberg, N.: Naturality in conformal field theory.
Nucl.\ Phys.\ B{\bf 313} 16-40 (1989)

\item{[19]} Ruelle, Ph.: Automorphisms of the affine SU(3) fusion rules.
Commun.\ Math.\ Phys.\ {\bf 160} 475-492 (1994)

\item{[20]} Ruelle, Ph. (private communications)

\item{[21]} Ruelle, Ph., Thiran, E., Weyers, J.: Modular invariants for
affine SU(3) theories at prime heights. Commun.\ Math.\ Phys.\ {\bf 133}
305-322 (1990)

\item{[22]} Ruelle, Ph., Thiran, E., Weyers, J.: Implications of an
arithmetical symmetry of the commutant for modular invariants. Nucl.\ Phys.\
B{\bf 402} 693-708 (1993)

\item{[23]} Schellekens, A.N.: Meromorphic $c=24$ conformal field theories.
Commun.\ Math.\ Phys.\ {\bf 153} 159-185 (1993)

\item{} Montague, P.: Orbifold constructions and the classification of
self-dual $c=24$ conformal field theories. (hep-th/9403088)

\item{[24]} Stanev, Y.: Classification of the local extensions of the
$SU(3)$ chiral current algebra.  Vienna preprint ESI 19 (May 1993)

\item{[25]} Stanev, Y.: Classification of the local extensions of
the $SU(2)\times SU(2)$ chiral current algebras. Vienna ESI preprint (April
1994).

\end